\documentclass[letterpaper,11pt]{article}
\usepackage[letterpaper,margin=1in]{geometry}
\usepackage{amsmath,amssymb,amsthm,mathtools,mathrsfs}
\usepackage{microtype}
\usepackage{thm-restate}
\usepackage{multicol}
\usepackage{flafter}
\usepackage[pdftex,breaklinks,colorlinks, pdfencoding=auto, psdextra, citecolor={BlueViolet}, linkcolor={Blue},urlcolor=Maroon]{hyperref}
\usepackage[colorinlistoftodos,prependcaption,textsize=tiny]{todonotes} \usepackage[shortlabels]{enumitem}
\usepackage[dvipsnames]{xcolor}
\usepackage{tabularx, booktabs, multirow, capt-of}
\usepackage[font+=smaller]{subcaption}
\usepackage{tikz}
\usetikzlibrary{calc, arrows.meta,positioning,fit,backgrounds,quotes}
\definecolor{pos}{HTML}{176B83}
\definecolor{neg}{HTML}{AE3F48}
\tikzset{
  empty vertex/.style = {circle, draw, fill = white, inner sep=1.5pt},
  filled vertex/.style = {circle,draw=black,fill=black!50,inner sep=1.5pt},
  positive/.style={draw=pos,thick},
  negative/.style={draw=neg, thick,densely dotted},
  cut/.style={RubineRed, dashed},
}

\allowdisplaybreaks
\newtheorem{theorem}{Theorem}[section]
\newtheorem{lemma}[theorem]{Lemma}
\newtheorem{corollary}[theorem]{Corollary}
\newtheorem{proposition}[theorem]{Proposition}
\newtheorem{definition}[theorem]{Definition}
\theoremstyle{remark}

\newcommand{\OPT}{\operatorname{OPT}}
\newcommand{\cost}{\operatorname{cost}}
\newcommand{\Gt}{\ensuremath{G\vee K_{t}}}

\newcommand{\E}{\mathbb E}
\newcommand{\Prob}{\mathbb P}
\newcommand{\ind}{\mathbf 1}
\newcommand{\calP}{\mathcal P}
\newcommand{\calQ}{\mathcal Q}
\newcommand{\calU}{\mathcal U}
\newcommand{\calT}{\mathcal T}
\newcommand{\calS}{\mathcal S}
\newcommand{\calW}{\mathcal W}
\newcommand{\binV}{\binom{V}{2}}

\title{Optimal Approximation for Constrained Correlation Clustering\\ via Repulsion}
\author{Yixin Cao\thanks{Department of Computing, Hong Kong Polytechnic University, Hong Kong, China.  \texttt{yixin.cao@polyu.edu.hk}, \texttt{ying-sophia-comp.xu@connect.polyu.hk}.} \and Ying Xu\footnotemark[1]}
\date{} 
\begin{document}
\pagenumbering{roman}
\maketitle

\begin{abstract}
The best known approximation factor for constrained correlation clustering stood at three for nearly two decades, until a recent improvement to below $16/7$.  We give a randomized polynomial-time $(2+\varepsilon)$-approximation using \emph{Repulsion}, a recursive operation that separates a metric ball from its complement.  Repulsion naturally respects hard constraints and admits a short potential-based analysis; as a warm-up, radius-$2/3$ Repulsion rounds the standard metric LP into a feasible partition of cost at most three times its objective.

The $(2+\varepsilon)$-approximation takes a two-stage approach, first removing must-links with arbitrarily small loss in the approximation ratio.  The second stage combines lossless, constraint-aware preclustering with the bounded sub-cluster LP framework of Cao, Cohen-Addad, Lee, Li, Newman, and Vogl (STOC 2024).  Repeated positive conditioning and sampling produce candidate sets, which radius-one Repulsion partitions into feasible clusters at controlled cost.
  
We also give an approximation-preserving reduction from vertex cover to cluster deletion, a special case with only cannot-links.  Apart from establishing a matching factor-two hardness threshold under the Unique Games Conjecture, it motivates our two-stage approach and Repulsion.
\end{abstract}

\setcounter{tocdepth}{2}
\begin{multicols}{2}
\small
\tableofcontents
\end{multicols}
\clearpage

\pagenumbering{arabic}

\section{Introduction}\label{sec:intro}

Correlation clustering is defined on a complete signed graph
$G=(V,E^+,E^-)$, where $E^+$ and $E^-$ partition $\binom V2$.  Positive pairs express a preference for being in the same cluster, while negative pairs express a preference for being separated.  The goal is to find a partition $\calP$ of $V$ minimizing disagreements: positive pairs between clusters and negative pairs within clusters.
The number of clusters is not prescribed.
Write $u\sim_{\calP}v$ when $u,v$ belong to the same cluster.
The cost of $\calP$ is
\[
 \cost_G(\calP)
 =\sum_{uv\in E^+}\ind[u\not\sim_{\calP}v]
  +\sum_{uv\in E^-}\ind[u\sim_{\calP}v].
\]
Equivalently, represent the instance by the unsigned graph
$(V,E^+)$, with negative pairs represented by nonedges.
The problem then asks for a minimum number of edge additions
and deletions that turn this graph into a disjoint union of
cliques; this is also known as the \emph{cluster editing} problem.

Constrained correlation clustering additionally specifies must-link pairs $M$ and cannot-link pairs $H$.
An instance is written $I=(V,E^+,E^-,M,H)$, and a partition
is feasible if every pair in $M$ lies within a cluster and
every pair in $H$ lies between clusters.
The objective remains $\cost_I(\calP)=\cost_G(\calP)$.  Must-links and cannot-links are also called friendly and hostile pairs, respectively.

Van Zuylen and Williamson~\cite{zuylen-09-deterministic-approximation} gave a $3$-approximation for this problem. Azizeddin et al.~\cite{azizeddin-26-constrained-correlation-clustering} obtained a randomized approximation with expected ratio below $16/7$, using the Sherali--Adams relaxation of the standard metric linear program (LP).  We improve the ratio arbitrarily close to two.

\begin{restatable}{theorem}{mainresult}\label{thm:main}
  For every fixed $\varepsilon>0$, there is a randomized algorithm that, on every feasible constrained correlation clustering instance $I$ with $n$ vertices, outputs in time $(n/\varepsilon)^{O(\varepsilon^{-2})}$ a feasible partition of expected cost at most $(2+\varepsilon)\OPT(I)$.  
\end{restatable}

We show a matching lower bound, assuming the Unique Games Conjecture (UGC).\footnote{The hardness results were obtained independently by Azizeddin et al.~\cite{azizeddin-26-constrained-correlation-clustering} and by Adriaens~\cite{adriaens-26-ckr-partitions}.  A recent preprint~\cite{openai-26-ugc} proposes a proof of the UGC.  If confirmed, it would make the factor-two NP-hardness unconditional.}

\begin{restatable}{theorem}{hardnessresult}\label{thm:p-time}
  Let $\varepsilon>0$ be a fixed constant.
  It is NP-hard to approximate constrained correlation clustering within a factor of~$\sqrt{2}-\varepsilon$.
  Assuming the UGC, it is NP-hard to approximate constrained correlation clustering within a factor of~$2-\varepsilon$.
\end{restatable}

\paragraph{Our techniques} Theorem~\ref{thm:p-time} applies to a special case with $M = \varnothing$ and $H = E^-$.
Equivalently, the \emph{cluster deletion} problem asks to delete the minimum number of edges to make the unsigned graph $(V, E^+)$ a disjoint union of cliques.
It is well known that an optimal cluster deletion solution does not need to preserve a maximum clique of the input graph, though greedily picking a largest clique of the remaining graph leads to a $2$-approximation~\cite{dessmark-07-edge-clique-partition}. (This procedure is not polynomial-time in general, since finding a maximum clique is NP-hard.)
For example, let $G$ be a graph whose vertex set can be partitioned into two cliques, $C = \{c_{1}, \ldots, c_{5}\}$ and $D = \{d_{1}, \ldots, d_{5}\}$,
and suppose that the only edges between $C$ and $D$ are $\{c_i d_j \mid 1\le i,j\le 3\}.$  Then $\{c_1,c_2,c_3,d_1,d_2,d_3\}$ is the unique maximum clique of~$G$, of size~$6$.  Nevertheless, the unique optimal cluster deletion solution deletes all nine edges between $C$ and $D$, thereby splitting this maximum clique.

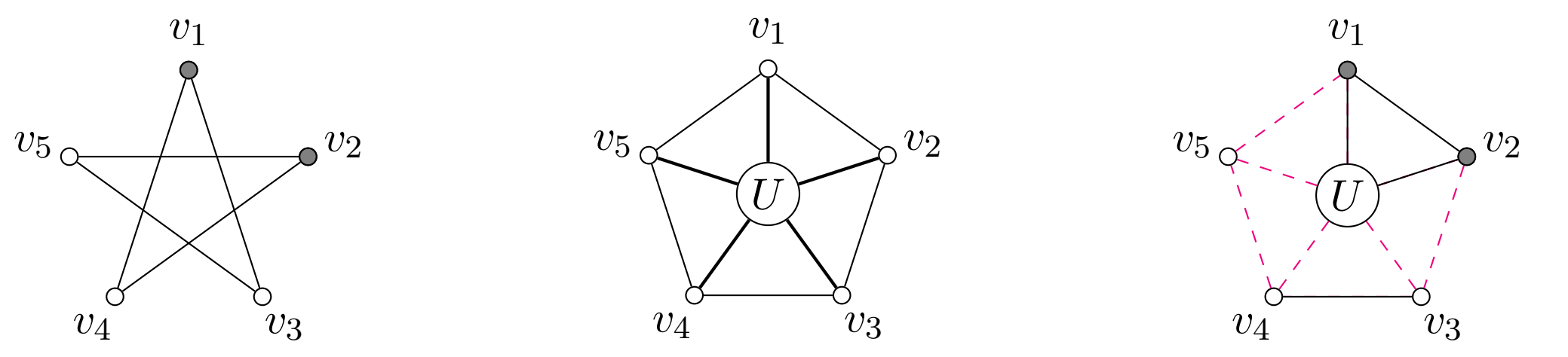
\begin{figure}[ht]
  \centering
  \begin{subfigure}[b]{0.3\linewidth}
    \centering
    \begin{tikzpicture}[scale=1.1]
      \def\n{5}
      \foreach \i in {0,..., \the\numexpr\n-1\relax} {
        \pgfmathsetmacro{\angle}{90 - \i * (360 / \n)}        
        \node[empty vertex] (v\i) at (\angle:1) {};
        \node at (\angle:1.3) {$v_{\number\numexpr\i+1\relax}$};
      }
      \foreach \i in {0,..., \the\numexpr\n-1\relax} {
        \pgfmathsetmacro{\j}{int(Mod(\i+2, \n))}
        \draw (v\i) edge (v\j);
      }
      \foreach \i in {0, 1}
      \node[filled vertex] at (v\i) {};
    \end{tikzpicture}
    \caption{$Q = C_{5}$.}
    \label{fig:vc-repulsion-a}
  \end{subfigure}
  \begin{subfigure}[b]{0.3\linewidth}
    \centering
    \begin{tikzpicture}[scale=1.1]
      \def\n{5}
      \node[empty vertex] (u) at (0, 0) {$U$};
      \foreach \i in {0,..., \the\numexpr\n-1\relax} {
        \pgfmathsetmacro{\angle}{90 - \i * (360 / \n)}        
        \node[empty vertex] (v\i) at (\angle:1) {};
        \node at (\angle:1.3) {$v_{\number\numexpr\i+1\relax}$};
        \draw[thick] (v\i) edge (u);
      }
      \foreach \i in {0,..., \the\numexpr\n-1\relax} {
        \pgfmathsetmacro{\j}{int(Mod(\i+1, \n))}
        \draw (v\i) edge (v\j);
      }
    \end{tikzpicture}
    \caption{$G$: join $U$ to $\overline Q$.}
    
  \end{subfigure}
  \begin{subfigure}[b]{0.3\linewidth}
    \centering
    \begin{tikzpicture}[scale=1.1]
      \def\n{5}
      \node[empty vertex] (u) at (0, 0) {$U$};
      \foreach \i in {0,..., \the\numexpr\n-1\relax} {
        \pgfmathsetmacro{\angle}{90 - \i * (360 / \n)}        
        \node[empty vertex] (v\i) at (\angle:1) {};
        \node at (\angle:1.3) {$v_{\number\numexpr\i+1\relax}$};
        \draw[cut] (v\i) edge (u);
      }
      \foreach \i in {1,..., \the\numexpr\n-1\relax} {
        \pgfmathsetmacro{\j}{int(Mod(\i+1, \n))}
        \draw[cut] (v\i) edge (v\j);
      }
      \draw (v0) -- (u) -- (v1) -- (v0) (v2) -- (v3);      
      \foreach \i in {0, 1}
      \node[filled vertex] at (v\i) {};
    \end{tikzpicture}
    \caption{Repulsion: $v_1$, then $v_2$}
    \label{fig:vc-repulsion-c}    
  \end{subfigure}
  \,
  \caption{The vertex cover reduction and Repulsion:
  (a) the graph $Q$; (b) the constructed graph $G$;
  (c) the clusters produced by Repulsion. Dashed edges are cut, and $U$ stays intact throughout.}
  \label{fig:vc-repulsion}
\end{figure}

The key observation behind our reduction is that this behavior changes after adding sufficiently many universal vertices.
If we extend a graph by adding a large number~$t$ of universal vertices, then every optimal cluster deletion solution in the resulting graph must place these universal vertices together with a maximum clique of the original graph.
Thus, from an optimal cluster deletion solution of the extended graph, one can recover a maximum clique of the original graph.
This gives a reduction from the clique problem, or equivalently from vertex cover in the complement graph, to cluster deletion; Figure~\ref{fig:vc-repulsion} illustrates the construction.
For sufficiently large $t$, the deletion cost closely tracks the cover size, allowing approximation guarantees to be transferred.

\begin{restatable}{theorem}{vctransfer}\label{thm:reduction}
  Let~$c>1$ be a constant.
  If there exists an $f(N)$-time $c$-approximation algorithm for cluster deletion on $N$-vertex graphs, then there exists an $ O(f(c'n^2)+n^4)$-time $c$-approximation algorithm for vertex cover on $n$-vertex graphs, where~$c'$ is a constant depending only on~$c$.
\end{restatable}

Theorem~\ref{thm:p-time} follows from Theorem~\ref{thm:reduction} and the vertex cover hardness results~\cite{khot-08-vc-ugc,khot-23-ugc}.  

This reduction suggests two algorithmic ideas. First, because
cannot-links alone suffice for factor-two hardness, we ask whether
must-links introduce an additional approximation loss. We show that
they do not, up to arbitrarily small slack.
Second, the connection with vertex cover suggests
Repulsion, a recursive separation operation that
naturally respects hard constraints.  Combining these ideas
with preclustering, the bounded sub-cluster LP, and sampling yields our algorithm.

\paragraph{Removing must-links.}

In an unsigned graph, two vertices are \emph{true twins} if they
are adjacent and have the same closed neighborhoods.
It is well known that every optimal solution to cluster editing or cluster deletion keeps true twins together.
We extend this preservation property to arbitrary pair costs with hard constraints.  The formal version is Theorem~\ref{thm:weighted-twins} in Section~\ref{sec:outline-mustlinks}.

\begin{theorem}[True-twin preservation, informal]
  \label{thm:weighted-twins-informal}
  Two vertices with identical external costs and hard constraints are kept together in every optimal feasible partition, provided they are not a cannot-link pair and separating them costs strictly more than keeping them together.  Moreover, every optimal solution of the metric LP and of the cluster LP assigns distance zero to the pair.
\end{theorem}

A recurring idea in our proofs is to identify, or create, groups of vertices that can be gathered into one cluster without increasing cost.  In particular, Theorem~\ref{thm:weighted-twins-informal} allows us to remove must-links by turning their components into
classes of weighted true twins.
This produces a probability-weighted
instance~\cite{bansal-04-correlation-clustering}. A constraint-preserving adaptation
of the random expansion of Chawla et al.~\cite{chawla-15-correlation-clustering}
then restores unweighted costs with arbitrarily small
approximation loss.  The main algorithm can therefore
focus on unweighted instances with only cannot-links.

When the input has no cannot-links, the expanded instance is ordinary correlation clustering.  Thus every approximation guarantee for ordinary correlation clustering transfers to must-link-only constrained correlation clustering up to arbitrary slack.

\begin{restatable}{theorem}{mustlinkonly}\label{thm:mustlink}
  For every fixed $\varepsilon>0$, there is a randomized polynomial-time $(1.3865+\varepsilon)$-approximation algorithm for constrained correlation clustering with must-link constraints only.  
\end{restatable}

Theorem~\ref{thm:mustlink} combines our reduction with the $(1.3865+\varepsilon)$-approximation of Garc{\'\i}a-Soriano and Schohn~\cite{garcia-soriano-26-1.387}.  It highlights a sharp dichotomy: must-links alone permit approximations below $\sqrt{2}$, whereas cannot-links immediately trigger the factor-two hardness barrier.
This transfer also has a potential use in hardness reductions for correlation clustering:
one may introduce must-links to simplify gadgets and then remove them without a constant-factor approximation loss.

\paragraph{Repulsion.}
Table~\ref{tab:vc-correspondences} in Section~\ref{sec:hardness} compares cluster-deletion methods with their vertex cover interpretations.
Although our reduction transfers approximation algorithms for cluster deletion to vertex cover, it also suggests how to translate an operation in the other direction.  Veldt~\cite{veldt-24-greedy-cover} showed that the complement of a random greedy maximal independent set is a $2$-approximate vertex cover in expectation.
In Figure~\ref{fig:vc-repulsion-a}, the greedy procedure selects $v_1$ and removes $\{v_1,v_3,v_4\}$, then selects $v_2$ and removes $\{v_2,v_5\}$.  It finds the independent set $S = \{v_1,v_2\}$ and the vertex cover $V(Q)\setminus S$.

Simulating this process on our constructed graph $G=\overline Q\vee U$, we select a vertex $v\in V(Q)$ and separate its remaining neighbors in $Q$, which are its non-neighbors in $G$.  Then $v$ remains with $U$.
Repeating this choice in the block containing $U$ finally leads to a clique of $\overline Q$, i.e., an independent set in $Q$.  
In Figure~\ref{fig:vc-repulsion-c}, processing $v_1$ separates $v_3,v_4$ from $U$, and processing $v_2$ in the remaining block separates $v_5$.  The resulting clusters are $U\cup\{v_1,v_2\}$, $\{v_3,v_4\}$, and $\{v_5\}$.
To solve cluster deletion, we may still need to repeat the operations on other blocks that are not yet cliques.  This leads to a linear-time randomized 2-approximation.

We call this clustering strategy \emph{Repulsion}, in contrast to \emph{Attraction}, which selects a block, finalizes it as a cluster, and continues on the remaining vertices.
Like Quicksort, Repulsion recursively processes both sides of
each split; neither side is immediately finalized as a cluster.
In the cluster deletion example, the chosen block is the closed neighborhood of a vertex in the remaining graph. More generally, we use strict balls in a feasible metric.
Repulsion is well suited to hard constraints: in a feasible metric, must-link endpoints have distance zero and hence identical distances to every center, so no metric-ball split separates them; terminating only when all pair distances in a block are less than one excludes every cannot-link.
In contrast, Attraction (Pivoting or CKR partitioning) with a ball radius greater than $1/2$ requires additional care to avoid placing cannot-link endpoints in the same cluster~\cite{charikar-05-approximate-clustering,chawla-15-correlation-clustering,adriaens-26-ckr-partitions}.

Repulsion on the metric LP gives a $3$-approximation, matching van Zuylen and Williamson~\cite{zuylen-09-deterministic-approximation}.

\begin{restatable}{theorem}{metricrepulsion}\label{thm:metric-repulsion}
Given any feasible solution $x$ of the standard metric LP for a constrained correlation clustering instance, randomized radius-$2/3$ Repulsion constructs a feasible partition in polynomial time with expected cost at most three times the objective value of $x$.
  The algorithm can be derandomized with the same guarantee.
\end{restatable}

\paragraph{Related work.}
Preclustering identifies atoms to be kept whole during rounding
and is central to recent correlation clustering
algorithms~\cite{cohen-addad-23-preclustering,cao-24-lp,cao-25-sublinear}.
Lossless preservation of dense sets also appears in parameterized
algorithms~\cite{cao-12-kernel-cluster-editing,cao-22-edge-modification}. Earlier approximations
combine atom preservation with restrictions on co-clustering
pairs~\cite{cohen-addad-23-preclustering,cao-24-lp}; ours accommodates cannot-links
while preserving every optimum.

Cao et al.~\cite{cao-24-lp} introduced the cluster LP and constructed
solutions of cost at most $(1+\varepsilon)\OPT$ in polynomial time
for fixed accuracy. Kalavas et al.~\cite{kalavas-25-constrained-correlation-clustering} showed that a
sufficiently accurate constrained cluster-LP solution would yield
a better-than-two approximation. Assuming the UGC and
$\mathrm{NP}\not\subseteq\mathrm{BPP}$, our hardness result
therefore rules out a direct extension with the same guarantee.
In particular, the sublinear-time solver of
Cao et al.~\cite{cao-25-sublinear} does not directly solve our
constrained relaxation. We instead round the bounded sub-cluster
LP directly.

Our sampling procedure adapts the size-and-pivot framework of Cao et al.~\cite[Section~4.2]{cao-24-lp}.  Lee et al.~\cite[Section~4.3]{lee-25-chromatic-correlation-clustering} extended size-and-pivot sampling to chromatic correlation clustering, and Fan et al.~\cite[Section~5]{fan-26-pseudometric-weighted-correlation-clustering} extended it to pseudometric-weighted correlation clustering. 

All known hardness results are through correlation clustering and cluster deletion.
Shamir et al.~\cite{shamir-04-graph-modification-problem} proved NP-hardness of approximation for cluster deletion within some constant
greater than one. Dessmark et al.~\cite{dessmark-07-edge-clique-partition}
gave the explicit threshold $881/880$; see also
\cite{adriaens-26-bad-triangle-transversals}.
Cao et al.~\cite{cao-24-lp} ruled out randomized $(24/23-\varepsilon)$-approximation for correlation clustering assuming $\mathrm{NP}\not\subseteq\mathrm{BPP}$. 

General weighted correlation clustering is approximation-equivalent, up to constant factors, to multicut~\cite{demaine-06-correlation-clustering}. Assuming the UGC, it is NP-hard to approximate either problem within any constant factor~\cite{chawla-06-hard-multicut}.
Large penalties can encode hard constraints on feasible instances, but not on infeasible ones.
Arbitrary two-weight instances, however, already capture general
weighted correlation clustering when the ratio between the two
weights is unbounded; see Section~\ref{sec:dis}.
Probability weights instead require separation and co-clustering
costs to sum to one on each pair~\cite{bansal-04-correlation-clustering,ailon-08-aggregating-inconsistent-information}.
Chawla et al.~\cite{chawla-15-correlation-clustering} reduced this
case to unweighted correlation clustering with arbitrarily small
approximation loss. We adapt their expansion to preserve cannot-links.

\section[Overview of techniques and main ideas]{Overview of techniques and main ideas}
\label{sec:outline}

This section presents the algorithmic framework and the main ideas behind the proof of Theorem~\ref{thm:main}.
The framework builds on Cao et al.~\cite{cao-24-lp}. We use their bounded sub-cluster LP and adapt their size-and-pivot sampling framework.
Our contributions are the treatment of hard constraints around this framework and the analysis that makes it work: must-links can be removed without a constant-factor loss; constraint-aware preclustering identifies indivisible atoms; sampling by positive conditioning produces candidates; and Repulsion partitions those candidates to respect cannot-links.
A summary of the algorithm is given at the end of this section (Figure~\ref{fig:outline}).
Detailed proofs appear in Sections~\ref{sec:reductions}--\ref{sec:repair} and the appendices.

Throughout this section, we assume that the input is feasible.  Infeasibility can be easily checked.

\subsection{Repulsion and the role of hard constraints}

Given a feasible metric $x$, Repulsion chooses a center $v$
in the current set $S$, separates the strict ball
\[
 A_v=\{u\in S:x_{uv}<\theta\}
\]
from its complement, and recurses on both parts.
Output a block as a cluster when all its pair distances are less than $\theta$.  Otherwise, choose a proper ball and split the block.
After the split, the chosen center $v$ is at distance less than $\theta$ from every vertex on its side.  This remains true in every later block containing~$v$, so it cannot define another proper ball.  Each split creates two nonempty parts, and there are at most $|V|-1$ splits.

\begin{figure}[ht]
  \centering
  \begingroup
\newcommand{\drawrepulsionstage}[1]{\begin{tikzpicture}[scale=1.1]
      \path[use as bounding box]
        (-1.7,-1.65) rectangle (1.7,1.65);
      \foreach \i in {1,...,7} {
        \pgfmathsetmacro{\angle}{90-(\i-1)*360/7}
        \coordinate (p\i) at (\angle:1.10);
        \coordinate (v\i) at (p\i);
        \coordinate (l\i) at (\angle:1.45);
      }
\foreach \i in {4,5} {
        \pgfmathsetmacro{\angle}{90-(\i-1)*360/7}
        \pgfmathsetmacro{\tangent}{\angle+90}
        \pgfmathsetmacro{\offset}{ifthenelse(\i==4,.16,-.16)}
        \pgfmathsetmacro{\back}{-\offset}
        \pgfmathsetmacro{\labeloffset}{1.5*\offset}
        \pgfmathsetmacro{\labelback}{-\labeloffset}
        \path (p\i) ++(\tangent:\offset) coordinate (v\i);
        \path (p\i) ++(\tangent:\back) coordinate (v\i p);
        \path (l\i) ++(\tangent:\labelback) coordinate (l\i p);
        \path (l\i) ++(\tangent:\labeloffset) coordinate (l\i);
      }

\draw[black!45] (v2) -- (v7);
      \foreach \u/\v in {1/2,1/7,3/4,3/4p,5/6,5p/6}
        \draw[thick] (v\u) -- (v\v);
      \draw[ForestGreen,very thick] (v4) -- (v4p);
      \draw[ForestGreen,very thick] (v5) -- (v5p);

\ifnum#1=0
        \foreach \u/\v in {1/3,1/6,2/4,2/4p,7/5,7/5p}
          \draw[black!45] (v\u) -- (v\v);
        \foreach \u/\v in {2/3,7/6}
          \draw[thick] (v\u) -- (v\v);
      \else
        \begin{pgfonlayer}{background}
          \foreach \u/\v in {1/3,1/6,2/3,2/4,2/4p,7/6,7/5,7/5p}
            \draw[cut] (v\u) -- (v\v);
        \end{pgfonlayer}
      \fi

\ifnum#1<2
        \foreach \u/\v in {3/5,3/5p,4/6,4p/6}
          \draw[black!45] (v\u) -- (v\v);
        \foreach \u/\v in {4/5,4/5p,4p/5,4p/5p}
          \draw[thick] (v\u) -- (v\v);
      \else
        \begin{pgfonlayer}{background}
          \foreach \u/\v in {3/5,3/5p,4/5,4/5p,4p/5,4p/5p,4/6,4p/6}
            \draw[cut] (v\u) -- (v\v);
        \end{pgfonlayer}
      \fi

\ifnum#1=0
        \path (v1) -- node[midway,above right,font=\scriptsize,
          fill=white,inner sep=.6pt] {$\frac13$} (v2);
        \path (v1) -- node[pos=.58,below left,font=\scriptsize,
          fill=white,inner sep=.6pt] {$\frac23$} (v3);
\path (v4) -- (v4p);
      \fi

      \foreach \i in {1,...,7} {
        \node[empty vertex] at (v\i) {};
        \node at (l\i) {$v_\i$};
      }
      \foreach \i in {4,5} {
        \node[empty vertex] at (v\i p) {};
        \node at (l\i p) {$v'_\i$};
      }
      \ifnum#1>0
        \node[filled vertex] at (v1) {};
      \fi
      \ifnum#1>1
        \node[filled vertex] at (v3) {};
      \fi
    \end{tikzpicture}}
  \begin{subfigure}[b]{.3\linewidth}
    \centering
    \drawrepulsionstage{0}
    \caption{The input and optimal metric.}
  \end{subfigure}
  \begin{subfigure}[b]{.3\linewidth}
    \centering
    \drawrepulsionstage{1}
    \caption{First Repulsion at $v_1$.}
  \end{subfigure}
  \begin{subfigure}[b]{.3\linewidth}
    \centering
    \drawrepulsionstage{2}
    \caption{Second Repulsion at $v_3$.}
  \end{subfigure}
  \endgroup
  \caption{Radius-$2/3$ Repulsion on a nine-vertex graph: it is obtained from $\overline{C_7}$ with each of $v_4,v_5$ replaced by a pair of true twins.
    All nonedges (negative) are cannot-links and have LP distance $1$.
    In the unique optimal metric LP solution, green twin edges have distance $0$, thick black edges have distance $1/3$, and thin gray edges have distance $2/3$.
    The first split separates $\{v_1,v_2, v_7\}$ from the other
    vertices. Within the latter block, the second split separates
    $\{v_3,v_4,v'_4\}$ from $\{v_5,v'_5,v_6\}$.
    Dashed edges are cut. These last two blocks are terminal;
    $\{v_7,v_1,v_2\}$ still requires one further split.}
  \label{fig:metric-repulsion-example}
\end{figure}
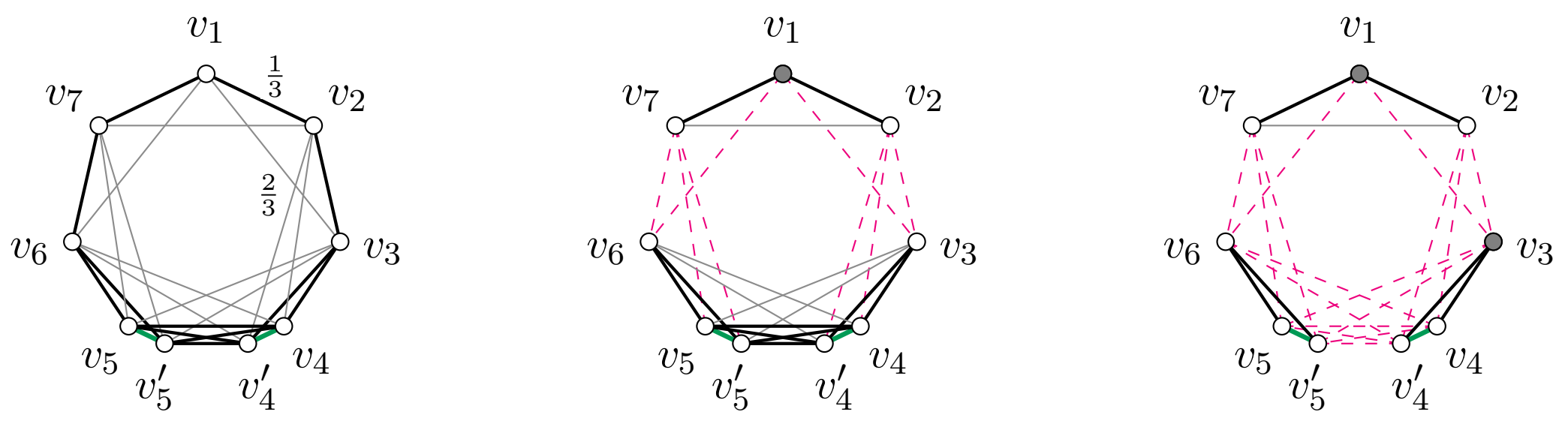

Let $I=(V,E^+,E^-,M,H)$.
For distinct vertices $u,v$, the variable $x_{uv}=x_{vu}$
represents separation. Set $x_{uu}=0$ by convention.
An integral partition assigns zero to pairs within a cluster
and one to pairs between clusters. The standard metric LP minimizes fractional disagreement cost:
\begin{equation}\label{eq:standard-metric-lp}
\begin{aligned}
 \min_x\quad
 &L(x)=\sum_{uv\in E^+}x_{uv} +\sum_{uv\in E^-}(1-x_{uv})\\
 \text{subject to}\quad
 &0\le x_{uv}\le1
       &&(u,v\in V,\ u\ne v),\\
 &x_{uv}\le x_{uw}+x_{wv}
       &&(u,v,w\in V),\\
 &x_{uv}=0
       &&(uv\in M),\\
 &x_{uv}=1
       &&(uv\in H).
\end{aligned}
\end{equation}
Figure~\ref{fig:metric-repulsion-example} illustrates the Repulsion behind Theorem~\ref{thm:metric-repulsion}, using an optimal solution of the metric LP.

Repulsion inherently respects the hard constraints.
The triangle inequalities imply that vertices at distance zero have identical distances to every other vertex.
If $u v$ is a must-link pair, then $x_{u v} = 0$ and hence $x_{u w}=x_{v w}$ for all $w\in V$.
Consequently, a metric ball contains either both $u$ and~$v$ or neither, and no split separates them.
A block whose pair distances are all less than one contains no cannot-link.
These feasibility properties hold for every radius $0<\theta\le1$.

\label{sec:pair-counts}
For $\sigma\in\{+,-\}$ and disjoint sets $A,B\subseteq V$, define
\[
 e^\sigma(A,B)=|E^\sigma\cap(A\times B)|,\qquad
 e^\sigma(A)=\left|E^\sigma\cap\binom A2\right|.
\]
We use $e^\sigma(v,A)=e^\sigma(\{v\},A\setminus\{v\})$,
so this notation also applies when $v\in A$.

\paragraph{Why radius $2/3$ gives factor three.}
Call a pair short if $x_{uv}<2/3$. Assign potential $3x_{uv}$
to short positive pairs, $1$ to short negative or non-short
positive pairs, and $0$ otherwise. The total is at most $3L(x)$.
Sum over centers the cut cost plus the change in potential.
In the graph of short pairs, the contributions group into
induced paths $u-v-w$, each contributing at most
\[
 2-3(x_{uv}+x_{vw})\le2-3x_{uw}\le0,
\]
since $x_{uw}\ge2/3$. Balls equal to the current set contribute
zero, so a uniform proper center has nonpositive expected excess;
choosing a minimum-excess center gives a deterministic version.
Terminal disagreement costs are covered by their potentials,
yielding the factor three (Appendix~\ref{sec:metric-repulsion}). 

\subsection[Replacing must-links with true twins]{Removing must-links without a constant-factor loss}
\label{sec:outline-mustlinks}

A must-link component must stay together.  As Azizeddin et
al.~\cite{azizeddin-26-constrained-correlation-clustering} observed, contracting it nevertheless creates a weighted
instance outside the unweighted problem setting.  We keep the vertex set intact and make each must-link component a class of true twins.

Here is the formal version of Theorem~\ref{thm:weighted-twins-informal}.
It is the only statement in the whole paper that allows arbitrary pair costs.
Let $w^+_{uv}$ and $w^-_{uv}$ denote the costs of separating and co-clustering $u,v$, respectively.
Only the differences between $w^+_{ij}$ and $w^-_{ij}$ affect the choice of a partition: subtracting the smaller one from both costs of each pair changes every partition cost by the same constant.

\begin{restatable}{theorem}{weightedtwins}\label{thm:weighted-twins}
  Consider a feasible instance $(V,E^+,E^-,M,H)$ with costs $w^+, w^-$.
  Every optimal feasible partition places distinct vertices $u,v$ together if $w^+_{uv}>w^-_{uv}$, $u v\not\in H$, and for every $t\notin\{u,v\}$:
\[
    u t\in M \iff v t\in M,\quad
    u t\in H \iff v t\in H,\quad
    w^+_{ut}-w^-_{ut}=w^+_{vt}-w^-_{vt}.
\]
Moreover, every optimal solution of the metric LP and of the cluster LP has $x_{uv}=0$.
\end{restatable}

\begin{figure}[htbp]
\centering
\begin{subfigure}[t]{.4\linewidth}
\centering
\begin{tikzpicture}[scale=.9,vertex/.append style={minimum size=5mm}]
\node[vertex] (a) at (0,0) {$a$};
\node[vertex] (b) at (-1,1.) {$b$};
\node[vertex] (c) at (-1,-1.) {$c$};
\node[vertex] (d) at (1.5, 0) {$d$};
\node[vertex] (e) at (-2.5, 0) {$e$};
\draw[positive] (a)--node[above right,font=\small] {$M$} (b);
\draw[positive] (a)--node[below right,font=\small] {$M$} (c);
\draw (a)--(d);
\draw (c)--(e);
\draw[negative] (b) to node[pos=.65, above,font=\small] {$H$} (e);
\end{tikzpicture}
\caption{Unweighted, $M = \{ab, ac\}$.}
\end{subfigure}
\begin{subfigure}[t]{.40\linewidth}
\centering
\begin{tikzpicture}[scale=.9,vertex/.append style={minimum size=5mm},
every edge quotes/.style={font=\scriptsize,fill=white,inner sep=1pt}]
\node[vertex] (a) at (0,0) {$a$};
\node[vertex] (b) at (-1,1.) {$b$};
\node[vertex] (c) at (-1,-1.) {$c$};
\node[vertex] (d) at (1.8, 0) {$d$};
\node[vertex] (e) at (-2.5, 0) {$e$};
\begin{pgfonlayer}{background}
  \draw[gray] (d) .. controls (1.8,2.2) and (-2.5,2.2) ..
    node[very near end, above,font=\scriptsize,fill=white,inner sep=1pt] {$0$} (e);
\end{pgfonlayer}
\draw (a)[positive] to["$1$"'] (b);
\draw[positive] (a)--(c);
\draw[positive] (b)--(c);
\draw (a) to["$1/3$"] (d);
\draw (b)--(d);
\draw (c)--(d);
\draw[negative] (b) to (e);
\draw[negative] (a)--(e);
\draw[negative] (c) to node[midway,below,font=\scriptsize] {$0$} (e);
\node[draw=pos,dashed,rounded corners,fit=(a)(b)(c),
      inner xsep=7pt] {};
\end{tikzpicture}
\caption{Probability-weighted.}
\end{subfigure}
\caption{Removing must-links by replacing the must-link component $\{a,b,c\}$ with a class of true twins.
  In (a), solid edges are positive, the pairs marked $M$ are must-links.  All other pairs are negative, including the cannot-link $b e$.  In (b), green pairs have separation cost $1$, black pairs have separation cost $1/3$, and the gray pair $de$ has separation cost $0$. Red dotted pairs have separation cost $0$ and are cannot-links.  The cannot-link is propagated to all pairs between $\{a,b,c\}$ and $e$, and the forced positive pair $ce$ is relabeled negative. The cost difference between the two instances is $2$: the negative pair $bc$ inside the must-link component and the positive pair $ce$.
}
\label{fig:mustlink-averaging}
\end{figure}
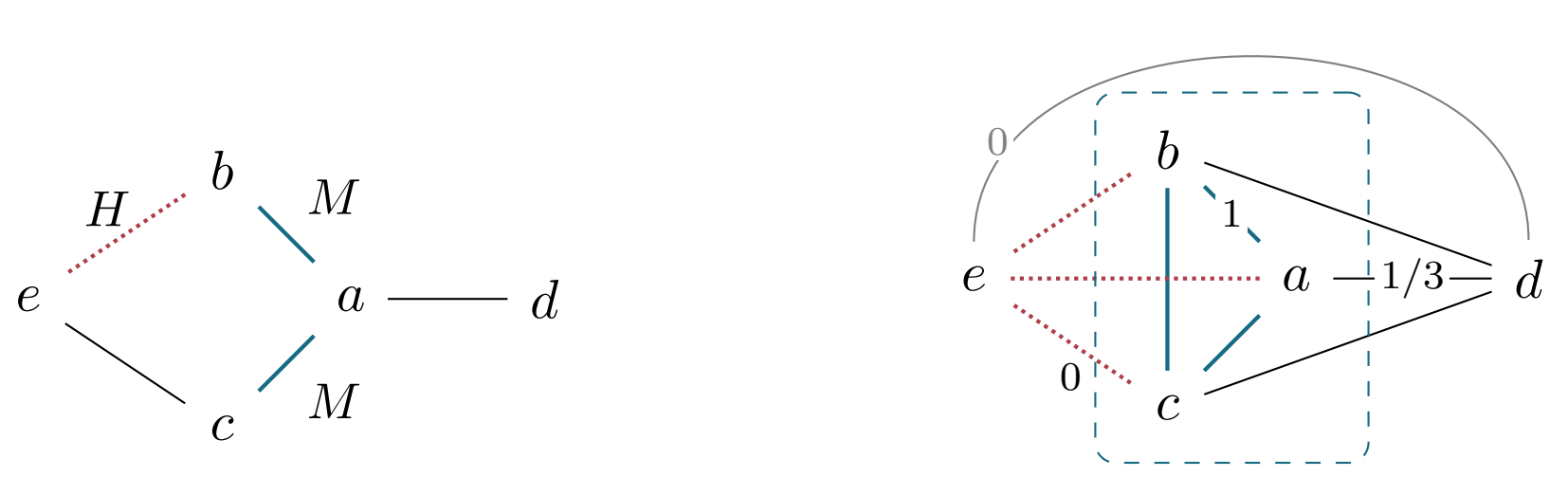

Our reductions use probability weights only.
We turn each must-link component into a set of true twins, assign unit weight to all its pairs, and replace its internal negative labels with positive labels.
If there is a cannot-link between two components, propagate it to all pairs between them and give these pairs separation cost zero.
For two compatible components $A, B$, assign the separation cost of every pair between $A$ and $B$ to be $\frac{e^+(A, B)}{|A||B|}$, the fraction of positive pairs between them.
This averaging maintains the total cost of separating $A$ and $B$.
Record the disagreement costs forced by the original constraints as a fixed additive cost.
Figure~\ref{fig:mustlink-averaging} illustrates both transformations.  
Theorem~\ref{thm:weighted-twins} ensures that every optimal partition preserves each must-link component, even after the explicit must-link constraints are removed.

The resulting costs have the probability-weighted form:
separation costs $p_{uv}\in[0,1]$ and co-clustering costs $1-p_{uv}$.
For correlation clustering, Chawla et al.~\cite[Section~6, Theorem~20]{chawla-15-correlation-clustering} reduced probability-weighted instances to the unweighted case with arbitrarily small loss in the approximation ratio.
Their random expansion also works for the constrained problem, preserving cannot-links by imposing them on all pairs between the corresponding groups of copies.
The two transformations remove must-links and keep the instance unweighted.

\begin{restatable}[Removing must-links]{theorem}{mustlinktransfer}
\label{prop:must-only-transfer}
Fix $\alpha\ge1$ and $\delta>0$.
A polynomial-time algorithm with expected approximation factor $\alpha$ for constrained correlation clustering with only cannot-links gives a randomized polynomial-time $(\alpha+\delta)$-approximation for general constrained correlation clustering.
\end{restatable}

We henceforth describe the algorithm on feasible unweighted instances with no must-links.

\subsection[Preclustering guarantee]{Preclustering: what can safely stay together?}\label{sec:outline-preclustering}

Ideally, we would identify all inclusion-wise maximal sets whose vertices are kept together in \emph{every} optimal partition.
Finding all these sets is computationally hard: even deciding whether two specified vertices are together in every optimal partition is coNP-hard (Section~\ref{sec:dis}).  

We therefore drop the maximality requirement and seek nontrivial sets, called \emph{atoms}, that every optimal feasible partition keeps together. Without a further requirement, one maximal set could contain several atoms. To avoid this, we also require that every optimal partition places distinct atoms in different clusters. Thus a maximal set may contain one atom together with other vertices, but cannot contain two atoms. We do not require every maximal set to contain a selected atom.

Intuitively, atoms have strong internal and weak external connectivity.  These properties suggest finding atoms inside the clusters of a constant-factor approximate partition: splitting a dense set or mixing it extensively with outsiders incurs disagreement cost.
Adapting the trimming construction of Cao et al.~\cite{cao-24-lp}, we mark all vertices free in clusters with large total incident disagreement, and trim vertices with large individual incident disagreement from the remaining clusters.

The additional obstacle is to handle cannot-links in the preservation proof.
To prove that a candidate is an atom, we assume that an optimal
feasible partition splits it and seek a cheaper feasible partition
that keeps it together. The exchange gathers its pieces into one
cluster, as in the true-twin argument.  An outside vertex already in that
cluster may have a cannot-link to the candidate and must then be
expelled, even if it has many positive neighbors there. We trim
the candidate further so that the saving from gathering its pieces
covers the cost of these expulsions.

We use a matching-based certificate to bound the cost of these expulsions: every outsider that could obstruct the exchange receives two distinct cannot-link witnesses inside the retained set, with no witness reused.
The additional vertices marked free to obtain this certificate are included in the degree charging argument. We use the following form of the $2$-expansion lemma~\cite{thomasse-10-kernel-fvs}.

\begin{restatable}[$2$-expansion]{lemma}{twoexpansion}\label{lem:two-expansion}
Given a bipartite graph with parts $L,R$, one can find in
polynomial time sets $Q\subseteq L$, $Z\subseteq R$, and an
edge set $M$ between them such that
\[
 N(Q)=Z,\qquad |L\setminus Q|\le2|R\setminus Z|.
\]
Every vertex of $Z$ is incident to exactly two edges of $M$,
and every vertex of $Q$ is incident to at most one.
The sets may be empty, and $Q$ contains every isolated vertex of $L$.
\end{restatable}

An atom contains at least two vertices.  A vertex not belonging to any selected atom is \emph{free}.
A \emph{unit} is an atom or a singleton consisting of a free vertex.
Let $\calU$ be the family of units and $B(v)$ the unit containing $v$.
Two distinct units are \emph{compatible} if no cannot-link joins them. 

Earlier preclustering frameworks further restricted which pairs could share a cluster: the vertices had to have comparable positive degrees and sufficiently many common positive neighbors. They then modified an optimal partition, allowing a small cost increase, to ensure that it respected these restrictions~\cite{cohen-addad-23-preclustering,cao-24-lp}. Our construction uses a simpler rule: every compatible pair between distinct units with a free endpoint remains allowed. Our sampling analysis needs a bound on the total positive degree of the free vertices, rather than on the number of allowed pairs. We therefore avoid the additional restrictions and preserve every optimum throughout preclustering.

For a vertex set $S$, write $e_S(v) =e^-(v, S)+e^+(v,V\setminus S)$ for the disagreements incident to $v$.
Figure~\ref{fig:preclustering} gives the algorithm.
The construction also bounds the total positive degree of free vertices.

\begin{figure}[tbp]
  \centering
  \begin{minipage}{.8\linewidth}
    \hrule\bigskip
    \small
    \begin{tabbing}
      AAA\=aaa\=aaa\=Aaa\=MMMMMAAAAAAAAAAAA\=A \kill
      1.\> use the deterministic version of Theorem~\ref{thm:metric-repulsion} to obtain $\calP$;
      \\
      2.\> \textbf{for each} cluster $S\in\calP$ \textbf{do}
      \\
      2.1.\>\> \textbf{if} $|S|=1$ or $\sum_{v\in S} e_S(v) \ge |S|^2/10000$ \textbf{then}
      \\
      \>\>\> mark every vertex of $S$ free and continue to the next cluster;
      \\
      2.2.\>\> $S' \leftarrow \{v\in S:e_S(v)\le |S|/100\}$;
      \\
      2.3.\>\> $W_S \leftarrow \{w\notin S:e^+(w,S)\ge |S|/4\}$;
      \\
      2.4.\>\> form the bipartite cannot-link graph between $S'$ and $W_S$;
      \\
      2.5.\>\> find $Q\subseteq S'$, $Z\subseteq W_S$, and $M$ as in Lemma~\ref{lem:two-expansion};
      \\
      2.6.\>\> add $Q$ as an atom, and mark every vertex of $S\setminus Q$ free;
      \\
      3.\> form the units from the atoms and singleton free vertices;
      \\
      4.\> add cannot-links between distinct atoms and propagate cannot-links between units;
      \\
      5.\> normalize: record and remove the fixed cost of positive cannot-link pairs;
    \end{tabbing}
    \hrule
  \end{minipage}
  \caption{The preclustering procedure. }
  \label{fig:preclustering}
\end{figure}

\begin{restatable}{theorem}{preclusteringresult}\label{thm:pre}
Let $I=(V,E^+,E^-,\varnothing,H)$ be an instance with
only cannot-links. There is an absolute constant
$C_{\mathrm{pre}}$ and a polynomial-time preclustering algorithm
that constructs units and a residual instance $I'$ on the same vertex set.
The following properties hold.
\begin{enumerate}[label=(\roman*)]
\item No cannot-link lies within a unit. Between distinct units $B,D$, either every pair is a cannot-link or none is. Every compatible pair of distinct units has a free endpoint.
\item Every optimal partition of the residual instance preserves each atom, and distinct atoms lie in different optimal clusters.
\item $\OPT(I')\le\OPT(I)$ and $\sum_{v \text{ is free}}e^+(v, V)\le C_{\mathrm{pre}}\OPT(I).$
\end{enumerate}
Every optimal partition of $I$ remains feasible for $I'$ and preserves the units.
Every feasible unit-preserving partition $\calP$ of $I'$ is
feasible for $I$ and satisfies
\[
 \cost_I(\calP)-\cost_{I'}(\calP)=\OPT(I)-\OPT(I').
\]
\end{restatable}

\paragraph{Why two witnesses suffice.}
Suppose an optimal partition $\calP^*$ splits an atom $Q$, and let $C\in \calP^*$ contain its largest piece $S=Q\cap C$ (breaking ties arbitrarily), leaving $R=Q\setminus C$.
Density estimates show that every outsider in $C$ incompatible
with $Q$ has enough positive neighbors in $Q$ to receive a
certificate. Let $X$ be these incompatible outsiders.
Since $C$ is feasible, both witnesses assigned to each vertex
of $X$ lie in $R$. No witness is reused, so $2|X|\le|R|.$
Reuniting $Q$ reduces the disagreement cost across $R$ and $S$ by approximately $|R||S|$, while separating $X$ from $S$ increases the cost by at most $|X||S|$. Since $2|X|\le |R|$, this increase offsets at most half of the $|R||S|$ term. The remaining cost saving is sufficient to cover the other cost increases under the low-disagreement bound on $Q$, so the exchange strictly decreases the total cost.

In the rest of this section, we refer to the residual instance $I'$ of Theorem~\ref{thm:pre} only.
All forced separations are encoded by cannot-links between units.

\subsection[The LP and its information]{The bounded sub-cluster LP and the information it provides}
\label{sec:outline-lp}

The metric LP has an integrality gap of at least $2$~\cite{chawla-15-correlation-clustering}, and how to round it to reach this bound remains open.
Pair distances alone do not specify the joint behavior needed after choosing a pivot. Even if two vertices each have conditional inclusion probability $1/2$, their joint inclusion probability could be zero or $1/2$. Independent draws would always give $1/4$. For example, take a three-vertex star
with center $c$ and leaves $u,v$, and average the partitions
$\{\{c,u\},\{v\}\}$ and $\{\{c,v\},\{u\}\}$.
Conditioned on a size-two cluster containing $c$, each leaf has inclusion probability $1/2$, but their joint inclusion probability is zero.  Independent draws instead include both with probability~$1/4$.

Cao et al.~\cite[Section~4.1]{cao-24-lp} introduced the bounded sub-cluster LP to record local joint information separately for each possible cluster size.  Solving the LP supplies fractional information for rounding.
Our use of the bounded sub-cluster LP differs from that of Cao et al.~\cite{cao-24-lp}. Their algorithm uses local inclusion information to construct a feasible solution to the full cluster LP without hard constraints, which then supports rounding with an approximation ratio below two. For our $(2 + \varepsilon)$-approximation, we bypass this intermediate construction.
Instead, we directly sample candidate sets from the bounded sub-cluster LP and use Repulsion to partition them into feasible clusters.

The key observation is that \emph{our cost analysis depends only on the vertex inclusion probabilities and the aggregate deviations of pairwise inclusion probabilities from their LP targets.}  The former are preserved exactly, while the latter can be controlled through seed conditioning. Moreover, our cost bound requires only the signed sum of sampling errors and the absolute error on positive pairs. This allows us to use the simpler, lossless preclustering, whose free-vertex degree bound suffices to control these errors.

We use the LP on whole units, still counting size in vertices.
Let $n=|V|$ and $[n]=\{1,\ldots,n\}$.
The LP variables describe the joint inclusion of units, separately
for each possible cluster size. Pairwise separation values are
derived from these variables.

\begin{definition}[$r$-bounded sub-cluster LP on units]
\label{def:moment-lp}
Fix an integer $r\ge3$. For $s\in[n]$ and
$\calT\subseteq\calU$ with $|\calT|\le r$, introduce a
variable $y_{s,\calT}$.
The LP is $\min L(x)$ subject to the following constraints
\begin{align}
&\sum_{s=1}^n y_{s,\{B\}}=1
 &&(B\in\calU),\label{lp:coverage}\\
 &\sum_{\calQ\subseteq\calT}(-1)^{|\calQ|}
       y_{s,\calS\cup\calQ}\ge0
 &&\left(\substack{s\in[n],\ \calS\cap\calT=\varnothing,\\
       \calS,\calT\subseteq\calU,\ |\calS|+|\calT|\le r}\right),
       \label{lp:SA}\\
 &\sum_{D\in\calU}|D|y_{s,\calT\cup\{D\}}
       =s\,y_{s,\calT}
 &&(s\in[n],\ |\calT|\le r-1),\label{lp:size}\\
 &y_{s,\{B,D\}}=0
 &&(s\in[n],\ B\times D\subseteq H), \label{lp:support}
\end{align}
where the separation values in the objective are defined from $y$ by
\begin{equation}\label{lp:pair}
 x_{BD}=1-\sum_{s=1}^n y_{s,\{B,D\}},\qquad
 x_{uv}=x_{B(u)B(v)}.
\end{equation}
\end{definition}

In an integral partition, $y_{s,\calT}$ counts the $s$-clusters
containing every unit in $\calT$; $y_{s,\varnothing}$ counts all
$s$-clusters and may exceed one.
The bounds
$0\le y_{s,\calS}\le1$ for nonempty $\calS$ are implicit.

Coverage~\eqref{lp:coverage} says that each unit belongs to one cluster.
The inclusion--exclusion inequalities~\eqref{lp:SA} imply
nonnegativity: taking $\calT=\varnothing$ gives $y_{s,\calS}\ge0$.
They also imply monotonicity: taking $\calT=\{D\}$ with
$D\notin\calS$ gives
$y_{s,\calS}-y_{s,\calS\cup\{D\}}\ge0$.
Thus requiring additional units cannot increase an inclusion value.
Consequently, for nonempty $\calS$ and $B\in\calS$,
\[
 0\le y_{s,\calS}\le y_{s,\{B\}}
 \le\sum_{i=1}^n y_{i,\{B\}}=1.
\]
The empty-set variables need not be at most one, but the size identity~\eqref{lp:size} and coverage~\eqref{lp:coverage} still give $0\le s\,y_{s,\varnothing} =\sum_{D\in\calU}|D|y_{s,\{D\}}\le n.$

The size identities~\eqref{lp:size} ensure that the label $s$ agrees with the
inclusion values in its layer. Fix a unit $B$ with
$y_{s,\{B\}}>0$ and divide the identity for $\calT=\{B\}$
by this value:
\[
 \sum_{D\in\calU}|D|
 \frac{y_{s,\{B,D\}}}{y_{s,\{B\}}}=s.
\]
Each ratio is the conditional inclusion value of $D$ given the size and pivot unit $B$. Hence the total expected number of included vertices is $s$, and the expected number outside $B$ is $s-|B|$. The same identity holds after conditioning on seeds and makes the seed probabilities sum to one. The final independent draws need not produce exactly $s$ vertices.

Equation~\eqref{lp:pair} connects these inclusion values with pairwise separation.  Through it the constraints \eqref{lp:SA}, \eqref{lp:coverage}, \eqref{lp:size}, and \eqref{lp:support} imply all constraints of the metric LP (Section~\ref{sec:lp-definition}).

For a concrete example of the probabilities these inequalities
provide, fix $s,B$ with $y_{s,\{B\}}>0$. For another unit $D$,
the values
\[
 \frac{y_{s,\{B,D\}}}{y_{s,\{B\}}},\qquad
 \frac{y_{s,\{B\}}-y_{s,\{B,D\}}}{y_{s,\{B\}}}
\]
represent including and excluding $D$, respectively, given that
$B$ is included. Both are nonnegative and sum to one.
More generally, the inequalities give a valid probability for
every inclusion and exclusion pattern on at most $r-1$ further
units. 
We do not impose the additional cluster-size gap used in~\cite[constraint (16)]{cao-24-lp}.
The local consistency constraints do not require the LP values to arise from a probability distribution over partitions of $V$.

\subsection{Sampling with seeds}

In an integral partition, the sampler simply selects a cluster
uniformly: choose its size $s$ with probability proportional to the
number of $s$-clusters, then choose a pivot uniformly among vertices
in those clusters. Conditioning on the pivot makes every unit's
inclusion value zero or one, so the independent draws return exactly
that cluster. The procedure below extends these choices to fractional
LP solutions without assuming a distribution on complete partitions.

Fix a rational optimal solution at an even level $r\ge6$ and let
$Z=\sum_s y_{s,\varnothing}$. For the chosen size $s$ and pivot $a$,
the eligible family is $\calW=\{D\in\calU\setminus\{B(a)\}: D\times B(a) \not \subseteq H\}$.
Initially
\[
 z_\calT=\frac{y_{s,\calT\cup\{B(a)\}}}{y_{s,\{B(a)\}}}
\]
is its conditional joint inclusion value.
A \emph{seed} is an eligible unit selected for inclusion; an update conditions
all remaining values on that choice.
The expected number of vertices outside the pivot unit under these local values is $t=s-|B(a)|$.
The size identities~\eqref{lp:size} make the pivot and seed probabilities sum to one. Figure~\ref{fig:sample} gives the sampling algorithm.
When $s=|B(a)|$, there is no outside inclusion mass, so Step~3 returns the pivot unit immediately.  The nontrivial case is $t > 0$.

\begin{figure}[tbp]
  \centering
  \begin{minipage}{.9\linewidth}
    \hrule\bigskip
    \small
    \begin{tabbing}
      Output: \= \kill
      Input: \> Optimal LP solution $y$ at even level $r\ge6$; units $\calU$; cannot-links $H$.
      \\
      Output: A candidate set $C \subseteq V$ consisting of whole units.
    \end{tabbing}

    \begin{tabbing}
      AAA\=Aaa\=aaa\=Aaa\=MMMMMAAAAAAAAAAAA\=A \kill
      1.\> sample $s\in\{1,\ldots,n\}$ with probability ${y_{s, \varnothing}}/Z$, with $Z = {\sum_{i=1}^n y_{i, \varnothing}}$;
      \\
      2.\> sample $a\in V$ with probability
      $\frac{y_{s, \{B(a)\}}}{s y_{s, \varnothing}}$;    \` {\footnotesize // $y_{s,\{B(a)\}}>0$ below.}
      \\
      3.\> \textbf{if} $|B(a)|= s$ \textbf{then return} $B(a)$;
      \\
      4.\> set $C=B(a)$; $t=s-|B(a)|$ and $\calW=\{D\in\calU\setminus\{B(a)\}:D\text{ is compatible with }B(a)\}$;
      \\
      5.\> \textbf{for} each $\calT\subseteq \calW$ with $|\calT|\le r-1$ \textbf{do} $\displaystyle z_\calT=\frac{y_{s, \calT\cup\{B(a)\}}}{y_{s, \{B(a)\}}}$;
      \\
      6.\> sample $K$ uniformly from $\{0,1,\ldots,r/2-1\}$;
      \\
      7.\> \textbf{for} $j=1,\ldots,K$ \textbf{do}
      \\
      7.1.\>\> sample a seed $D\in \calW$ with probability ${|D|z_{\{D\}}}/{t}$;    \` {\footnotesize // repetition allowed.}
      \\
      7.2.\>\> \textbf{for} each $\calT\subseteq \calW$ with $|\calT|\le r-1-j$ \textbf{do}  $ z_\calT^{\mathrm{new}} = {z_{\calT\cup\{D\}}}/{z_{\{D\}}}$;
      \\
      7.3.\>\> $z \leftarrow z^{\mathrm{new}}$;
      \\
      8.\> \textbf{for} every $D\in \calW$ \textbf{do} add all vertices of $D$ to $C$ with probability $z_{\{D\}}$ (independently);
      \\
      9.\> \textbf{return} $C$.\` {\footnotesize // possibly $|C|\ne s$.}
    \end{tabbing}
    \hrule
  \end{minipage}
  \caption{Sampling one candidate by positive conditioning.}
  \label{fig:sample}
\end{figure}

Step~7 reduces the error caused by the final independent draws.
For two distinct units $B,D$, those draws use the product
$z_{\{B\}}z_{\{D\}}$ instead of the joint value $z_{\{B,D\}}$.
Selecting a seed conditions the inclusion values on its inclusion.
Repeating this operation for a random number of steps controls
the aggregate discrepancy while preserving inclusion values in
expectation.  Every selected seed has inclusion probability one and belongs to the returned candidate.
Selecting a seed also sets the inclusion value of every unit at distance one from it to zero. Thus, repulsion against selected seeds is already implicit in conditioning.

Seeds are sampled with replacement. 
Selecting an already chosen seed leaves all retained table entries unchanged: its inclusion value is already one, and the table is already conditioned on it. The repeated selection still counts as one seed step. In Step~7.2, every new
entry is computed from the old entries. Step~7.3 replaces the
entries only after all these computations are complete; overwriting
entries during Step~7.2 could change a later numerator or denominator.

If the LP solution is induced by an integral partition, the
sampler always returns one of its clusters, with exactly $s$
vertices.  For a fractional solution, the independent choices
in Step~8 can produce $|C|\ne s$.

\paragraph{The variance-drop identity.}
Fix the size and pivot with $t>0$, and put
$W=\bigcup_{U\in\calW}U$. After $k$ updates, write
$z_i=z_{\{B(i)\}}$ and $z_{ij}=z_{\{B(i),B(j)\}}$ for
$i,j\in W$. Let $\mathcal H_k$ record the selected seeds and \[
 c_{ij}=z_{ij}-z_i z_j,\qquad
 \Phi_k=\sum_{i\in W}z_i(1-z_i),\qquad
 \mathcal E_k=\sum_{i,j\in W: z_j\ne 0}\frac{c_{ij}^2}{z_j}.
\]
When $z_j=0$, the inclusion inequalities~\eqref{lp:SA} also give $z_{ij}=0$. The sums include ordered pairs and
vertices in the same unit. The size identity gives
$\sum_{i\in W}z_i=t$, so $0\le\Phi_k\le t$.
Selecting a seed unit is equivalent to selecting a vertex $j$
with probability $z_j/t$ and conditioning on its unit. The update
changes $z_i$ by $c_{ij}/z_j$, and the size identities~\eqref{lp:size} imply that
its conditional expectation is zero. Expanding the variance gives
\[
 \Phi_k-\E[\Phi_{k+1}\mid\mathcal H_k]
 =\frac{\mathcal E_k}{t}.
\]
Consider $r/2$ possible updates. When $K=k$, the sampler stops
before update $k+1$. The stopping index $K$ is uniform in
$\{0,\ldots,r/2-1\}$.
Taking expectations and summing the identity gives
\[
 \E[\mathcal E_K\mid s,a]
 =\frac{2t}{r}\bigl(\Phi_0-\E[\Phi_{r/2}\mid s,a]\bigr)
 \le\frac{2t^2}{r}\le\frac{2s^2}{r}.
\]
This is a bound on the aggregate discrepancy between joint
inclusion values and products of individual inclusion values,
which is precisely the discrepancy introduced by Step~8.
The variance does not need to decrease for each seed outcome.
We show that level $r=O(\varepsilon^{-2})$ suffices.

Our sampler adapts the size-and-pivot framework of Cao et al.~\cite{cao-24-lp}. Our seed analysis bounds sampling error directly using LP inclusion values and free-vertex degrees. It thus works without the cluster-size gap and small admissible-pair set used in their analysis of Raghavendra--Tan rounding.

\subsection[Final Repulsion and cost]{Repulsion restores feasibility at controlled cost}
\label{sec:outline-cost}
The final independent step can select two incompatible units even when their joint
LP inclusion value is zero. We therefore intersect each candidate
with the unprocessed vertices and apply radius-one Repulsion to this
intersection. We output its feasible clusters and remove its vertices.
The intersection must precede Repulsion because the cost accounting
uses the first round that processes either endpoint of a pair.

\paragraph{Where the factor two comes from.}
Fix a pair $uv$, put $x=x_{uv}$, and let $q$ be its joint
inclusion probability in one candidate. Each endpoint has marginal
$1/Z$. At the first round selecting either endpoint, the conditional
probabilities of selecting exactly one and selecting both are
\[
 \frac{2-2qZ}{2-qZ},\qquad \frac{qZ}{2-qZ},
\]
respectively.  Exactly one endpoint selected separates the pair.
When both are selected, the pair enters the potential bound for
Repulsion. Let $a=\ind[uv\in E^+]$, and put $\xi=qZ$. The resulting accounting contribution is
\[
 F_a(x,\xi)=\frac{2a(1-\xi)+(1-a+a b(x))\xi}{2-\xi},
 \qquad b(x)=\min\left\{2,\frac{2x^2}{1-x}\right\},
\]
with $b(1)=2$. Lemma~\ref{lem:firsthit} bounds the expected total cost by the sum of these pair contributions.
If joint sampling were exact, $\xi=1-x$, then
\[
 \begin{aligned}
 F_1(x,1-x)&=\min\left\{2x,\frac{2}{1+x}\right\}\le2x,\\
 F_0(x,1-x)&=\frac{1-x}{1+x}\le1-x.
 \end{aligned}
\]
Summing gives at most twice the positive part plus once the
negative part of the LP objective, and hence at most twice its
total value. Hard pairs have zero ideal joint inclusion and no
cost after normalization. These exact-probability calculations
explain the leading two.  The remaining task is to bound the
cost of replacing ideal joint probabilities by sampled ones.

Lemma~\ref{lem:discrepancy} bounds both the squared pair errors
and the signed error in the cost expression by $1/r$ times the
LP co-clustering mass on pairs in different units, up to an
absolute constant. Lemma~\ref{lem:weighted-error}
bounds this mass and the positive cross-unit pair count by the
free-vertex degree budget plus $L$, hence by a constant times
$\OPT(I)$. Cauchy--Schwarz gives absolute positive-pair error
$O(\OPT(I)/\sqrt r)$. Thus preclustering converts additive
sampling error into relative error without dependence on the
largest unit size.

Combining Lemma~\ref{lem:rounding-bound} with the normalization
identity in Theorem~\ref{thm:pre} gives
\[
 \E[\cost_I(\text{infinite-round algorithm})]
 \le2\OPT(I)+c\OPT(I)\left(\frac1r+\frac1{\sqrt r}\right)
\]
for an absolute constant $c$. Choosing $r$ as a sufficiently large
constant multiple of $\varepsilon^{-2}$ makes this overhead small.
Lemma~\ref{lem:truncate} bounds the additional loss from stopping
after finitely many rounds and outputting remaining units separately.
Section~\ref{sec:completion} combines the sampling, truncation, and
reduction bounds to prove Theorem~\ref{thm:main} with running time
$(n/\varepsilon)^{O(\varepsilon^{-2})}$.

\begin{figure}[tbp]
  \centering
  \begin{minipage}{.8\linewidth}
    \hrule\bigskip
    \small
    \begin{tabbing}
      AAA\=Aaa\=aaa\=Aaa\=MMMMMAAAAAAAAAAAA\=A \kill
      1.\> reduce to an instance with only cannot-links;
      \\
      2.\> precluster into units and normalize forced separation costs;
      \\
      3.\> solve the bounded sub-cluster LP at an even level $r=\Theta(\varepsilon^{-2})$;
      \\
      4.\> $\calP \leftarrow \varnothing$;
      \\
      5.\> \textbf{repeat} $T$ times:  \` {\footnotesize // $T = O(n \log(n/\varepsilon))$, specified in Section~\ref{sec:completion}.} \\
      5.1.\>\> sample a candidate $C$; set $D = C\setminus \bigcup_{B\in\calP}B$;
      \\
      5.2.\>\> partition $D$ by Repulsion;
      \\
      5.3.\>\> add the resulting clusters to $\calP$;
      \\
      6.\> add each remaining unit to $\calP$ as a separate cluster;
      \\
      7.\> project $\calP$ back to the input and \textbf{return} the resulting partition.
    \end{tabbing}
    \hrule
  \end{minipage}
  \caption{Outline of the approximation algorithm.}
  \label{fig:outline}
\end{figure}

\paragraph{Organization.}
Sections~\ref{sec:reductions}--\ref{sec:atoms} prove the reductions
and preclustering guarantee. Sections~\ref{sec:lp}--\ref{sec:repair}
give the sampling analysis, Repulsion bound, and main proof.
Section~\ref{sec:hardness} proves approximation hardness. Section~\ref{sec:dis} proves the hardness of finding all sets preserved by every optimum, discusses derandomization, and considers two-weight instances. The appendices give the proofs of metric-LP Repulsion and the probability-weighted expansion.

\section[Reduction to cannot-links only]{Reduction to instances with only cannot-links}
\label{sec:reductions}

We reduce general constrained correlation clustering to an
unweighted instance with only cannot-links. The proof of
Theorem~\ref{prop:must-only-transfer} has two steps. 
We replace must-link components with weighted true-twin classes, and then remove fractional weights by a random expansion that preserves cannot-links between groups of copies.
The intermediate instance uses the standard probability-weighted
formulation~\cite{bansal-04-correlation-clustering,ailon-08-aggregating-inconsistent-information}: each pair $uv$ has separation cost
$p_{uv}\in[0,1]$ and co-clustering cost $1-p_{uv}$.
For a partition $\calP$, write
\[
 \cost_p(\calP)
 =\sum_{u<v}
 \left[
 p_{uv}\ind[u\not\sim_{\calP}v]
 +(1-p_{uv})\ind[u\sim_{\calP}v]
 \right].
\]
Only the intermediate instance is weighted; the input and the reduced instance used by the main algorithm are unweighted.

We use the following normalization at several points in the algorithm.

\begin{definition}[Normalization]\label{def:normalization}
  Relabel positive cannot-link pairs as negative and record their forced cost.  Then remove vertices with no positive neighbors.
  An unweighted instance with only cannot-links is \emph{normalized} if every cannot-link pair is negative and every vertex has at least one positive neighbor.
\end{definition}

Every feasible partition of the remaining instance extends by restoring the removed vertices as singletons, adding only the recorded cost. The optimal value drops by the same nonnegative constant, so approximation ratios are preserved. Normalization takes linear time in an explicit
representation of all pairs.

\subsection[True-twin replacement]{Removing must-links by true-twin replacement}
\label{sec:mustlink-twins}

We first prove the true-twin preservation property, which allows arbitrary real separation and co-clustering costs.
For the LP claims, both relaxations minimize the weighted objective
\[
 \sum_{uv\in\binom V2}
 \bigl[w^+_{uv}x_{uv}+w^-_{uv}(1-x_{uv})\bigr].
\]
The metric LP uses the constraints in~\eqref{eq:standard-metric-lp}.
For the full cluster LP, let $\mathcal C$ be the family of nonempty
sets $C\subseteq V$ containing no cannot-link pair and containing
either both endpoints or neither endpoint of every must-link.
Introduce a variable $\lambda_C\ge0$ for each $C\in\mathcal C$,
and impose
\[
 \sum_{C\in\mathcal C:\,v\in C}\lambda_C=1\qquad(v\in V),
 \qquad
 x_{uv}=1-\sum_{C\in\mathcal C:\,u,v\in C}\lambda_C
 \qquad(u\ne v).
\]
Thus $\lambda_C$ is a fractional weight on a feasible cluster set,
and every vertex has total coverage one.

\weightedtwins*
\begin{proof}
If $uv\in M$, every feasible partition keeps $u,v$ together and
every feasible LP solution has $x_{uv}=0$.
If $ut\in M$ for some $t\notin\{u,v\}$, the assumed symmetry
also gives $vt\in M$. Feasibility then forces $u,v$ together;
the metric triangle inequality gives
$x_{uv}\le x_{ut}+x_{tv}=0$, and every feasible cluster-LP
cluster containing either vertex must contain both.
We may therefore assume that neither $u$ nor $v$ is incident
to a must-link.

Put $c_{ij}=w^+_{ij}-w^-_{ij}$. The objective equals
\[
 \sum_{i<j}w^-_{ij}
 +\sum_{i<j}c_{ij}\ind[i,j\text{ separated}].
\]
The first sum is independent of the partition.
Suppose $u,v$ belong to different clusters. Moving $u$ into
$v$'s cluster or moving $v$ into $u$'s cluster is feasible
because their external cannot-link relations agree.
The two cost changes satisfy
\[
 \Delta_u+\Delta_v=-2c_{uv}.
\]
Indeed, their contributions involving any other vertex cancel,
while each move removes the separation cost $c_{uv}$.
Since $c_{uv}>0$, at least one move strictly improves the
partition. Comparing the two cost changes also gives a
polynomial-time improving move whenever the pair is separated.
This argument does not require either pair cost to be nonnegative.

For the metric LP, clone $v$ onto $u$ by setting $x_{uw}=x_{vw}$
for $w\notin\{u,v\}$ and $x_{uv}=0$, or clone $u$ onto $v$.
Each operation preserves bounds, triangle inequalities, and
cannot-link distances. Their objective changes sum to
$-2c_{uv}x_{uv}$.
For the cluster LP, the analogous operations remove one vertex
from every cluster set and insert it exactly in sets containing
the other vertex, combining equal sets and discarding empty sets.
They preserve coverage one at every vertex, nonnegativity, and
cannot-link feasibility. Their pair distances have the same
cloning behavior, and hence the same objective identity.
If the solution is optimal, neither cloning operation decreases
its objective. Hence their changes sum to
$-2c_{uv}x_{uv}\ge0$. Since $c_{uv}>0$, this implies $x_{uv}=0$
and proves both LP claims.
\end{proof}

Given a feasible instance $I=(V,E^+,E^-,M,H)$, construct $J$
as follows. Let $B_1,\ldots,B_h$ be the connected components
of $(V,M)$. Every feasible partition must preserve each $B_i$.
The input is infeasible exactly when a cannot-link has both
endpoints in one component.

Call two distinct components incompatible if a cannot-link
joins them. Impose cannot-links on every pair between
incompatible components. The fixed disagreement cost forced
by the input constraints is
\begin{equation}\label{eq:mustlink-forced-cost}
 \sum_i e^-(B_i)
 +\sum_{\substack{i<j\\(B_i\times B_j)\cap H\ne\varnothing}}
       e^+(B_i,B_j).
\end{equation}

Within each component set $p_{uv}=1$.
Between incompatible components set $p_{uv}=0$.
For compatible distinct components, set
\[
 p_{uv}
 =\frac{e^+(B_i,B_j)}{|B_i||B_j|}
 \qquad(u\in B_i,\ v\in B_j).
\]
Remove the must-link constraints to obtain $J$.

The next lemma proves the optimal-cost relation and the
projection guarantee for this construction.

\begin{lemma}\label{lem:averaging}
Given a feasible complete unweighted constrained correlation
clustering instance $I=(V,E^+,E^-,M,H)$, one can construct in
polynomial time a probability-weighted instance $J$ on the
same vertex set, with only cannot-links and
$\OPT(J)\le\OPT(I)$.
Every feasible partition $\calQ$ of $J$ can be projected
deterministically in polynomial time to a feasible partition
$\calP$ of $I$ satisfying
\begin{equation}\label{eq:M-projection}
 \cost_I(\calP)-\cost_J(\calQ)
 \le\OPT(I)-\OPT(J).
\end{equation}
\end{lemma}
\begin{proof}
Every feasible partition of $I$ preserves the components $B_i$.
For any such partition $\calP$, the construction gives
$\cost_I(\calP)-\cost_J(\calP)$ equal to
\eqref{eq:mustlink-forced-cost}.
Conversely, Theorem~\ref{thm:weighted-twins} implies that every
optimum of $J$ preserves the true-twin classes $B_i$, so it is
also feasible for $I$.
Applying this cost identity to the optima shows that
$\OPT(I)-\OPT(J)$ is also given by
\eqref{eq:mustlink-forced-cost}. In particular,
$\OPT(J)\le\OPT(I)$.

For the projection, apply the constructive true-twin argument
from the proof of Theorem~\ref{thm:weighted-twins} to gather
each class $B_i$ deterministically, preserving feasibility and
without increasing $\cost_J$. The resulting partition $\calP$
is feasible for $I$ and satisfies
$\cost_J(\calP)\le\cost_J(\calQ)$.
Together with the cost identity above, this proves \eqref{eq:M-projection}.

Both the construction and the projection take polynomial time.
Within each $B_i$, the two twin moves have cost changes summing to $-2$; choosing the better move decreases cost by at least one, so at most $\binom n2$ moves suffice.
\end{proof}

If $\calQ$ splits any component, gathering strictly decreases
its internal separation cost, so \eqref{eq:M-projection} is strict.
If $\calQ$ already preserves every component, we may take $\calP=\calQ$ and obtain equality.
In particular, this is the case when $\calQ$ is an optimal partition of $J$ by Theorem~\ref{thm:weighted-twins}.

\subsection{Reduction to unweighted instances}
\label{sec:must-only}

We next remove the fractional costs while preserving cannot-links.
The unconstrained expansion is due to Chawla et al.~\cite[Section~6, Theorem~20]{chawla-15-correlation-clustering}.
For the sake of completeness, we give a proof in Appendix~\ref{app:probability-transfer}.
Our extension imposes all cross-group cannot-links between copies.
The expanded instance has polynomial size in the original
vertex count and $1/\delta$.

\begin{restatable}[\cite{chawla-15-correlation-clustering}]{theorem}{probabilitytransfer}
\label{prop:constraint-transfer}
Fix constants $\alpha\ge1$ and $\delta>0$.
Suppose complete unweighted constrained correlation clustering with only cannot-links admits
a polynomial-time algorithm that always returns a feasible
partition and has expected approximation factor $\alpha$.
Then probability-weighted constrained correlation clustering with only cannot-links admits
a randomized polynomial-time algorithm that always returns
a feasible partition and has expected approximation factor
$\alpha+\delta$.
\end{restatable}

The reduction replaces each vertex $v$ by a positive clique $V_v$ of copies.
Between copies of distinct vertices $u,v$,
it independently samples positive labels with probability
$p_{uv}$ and negative labels otherwise.
For every cannot-link $uv$, it imposes cannot-links on
\emph{all} pairs in $V_u\times V_v$.
No must-links are imposed within the groups of copies.

Every feasible original partition lifts to a feasible partition
of the expanded instance by keeping each group together.
Conversely, from any feasible expanded partition, independently
choose a uniform representative from each group and assign
the original vertex its representative's cluster label.
Every original cannot-link is respected because every possible
pair of its representatives is forbidden in the expanded instance.

The concentration bound must hold for every partition at once, because the algorithm chooses its partition after seeing the sampled labels.
The reduction may solve an instance directly instead of expanding it.

\mustlinktransfer*
\begin{proof}
Apply Lemma~\ref{lem:averaging} to obtain a probability-weighted
instance $J$ with only cannot-links.
Apply Theorem~\ref{prop:constraint-transfer} to $J$, obtaining
a feasible partition $\calQ$, and project it back to a feasible
partition $\calP$ of $I$ using Lemma~\ref{lem:averaging}.
By \eqref{eq:M-projection},
\[
 \begin{aligned}
 \E[\cost_I(\calP)]
 &\le \OPT(I)-\OPT(J)+\E[\cost_J(\calQ)]\\
 &\le \OPT(I)+(\alpha+\delta-1)\OPT(J)\\
 &\le(\alpha+\delta)\OPT(I).
 \end{aligned}
\]
The last inequality uses $\OPT(J)\le\OPT(I)$ and
$\alpha+\delta\ge1$.
Both transformations preserve feasibility and take polynomial
time, proving the theorem.
\end{proof}

When the input has no cannot-links, neither true-twin replacement nor expansion introduces any. The expanded instance is then ordinary complete unweighted correlation clustering.

\mustlinkonly*
\begin{proof}
Apply Theorem~\ref{prop:must-only-transfer} with slack
$\delta=\varepsilon/2$ and the correlation clustering algorithm
of Garc\'ia-Soriano and Schohn~\cite{garcia-soriano-26-1.387} with accuracy
$\varepsilon/2$. With no original cannot-links, neither true-twin replacement nor expansion introduces any, so this algorithm applies to the
expanded instance. The resulting expected ratio is at most
$1.3865+\varepsilon$, and the projection preserves every must-link.
\end{proof}

For the main $(2+\varepsilon)$ guarantee, use the unweighted
cannot-link algorithm with approximation factor
$\alpha=2+\varepsilon/2$ and allow $\delta=\varepsilon/2$
in Theorem~\ref{prop:must-only-transfer}.
Thus preclustering and rounding need only handle an unweighted
instance with no must-links.
The probability weights used in true-twin replacement do not enter the main algorithm.

If the expanded instance has
$N=\operatorname{poly}(n,1/\varepsilon)$ vertices and the core
algorithm runs in $N^{O(\varepsilon^{-2})}$ time, the resulting
running time is $(n/\varepsilon)^{O(\varepsilon^{-2})}$.
Vertex counts and recorded reductions in the analysis of the
core algorithm refer to this expanded instance.
After clustering it, we project from copies to the averaged
instance and then from must-link components to the original input.

\section{Constraint-aware preclustering}\label{sec:atoms}

We apply preclustering to the feasible unweighted instance
$I=(V,E^+,E^-,\varnothing,H)$ obtained from
the normalization in Definition~\ref{def:normalization}. An \emph{atom} is a set of at least
two vertices that every optimal feasible partition keeps
together. We do not require atoms to be maximal, but our
construction also ensures that every optimum places distinct
selected atoms in different clusters.
A vertex outside the selected atoms is \emph{free}; let $F$
be the set of free vertices. A \emph{unit} is an atom or a
singleton free vertex. Write $\calU$ for the family of units
and $B(v)$ for the unit containing $v$. Two distinct units
are \emph{compatible} if no cannot-link joins them.

Our construction adapts the trimming procedure of
Cao et al.~\cite[Section~3.1]{cao-24-lp} to handle cannot-links.
We start from the feasible $3$-approximate partition supplied by
Theorem~\ref{thm:metric-repulsion}. We first mark all vertices
free in singleton clusters and in clusters with large total
incident disagreement. From each remaining cluster, we trim
vertices with large individual incident disagreement: negative
pairs inside the cluster or positive pairs leaving it.
The candidate sets obtained this way have strong internal
connectivity and weak external connectivity.

We need one more trimming step. To prove a candidate is indeed
an atom, we show that the cost of a solution splitting it can
be decreased by gathering its pieces into one cluster.
To ensure that this exchange respects cannot-links, we trim
the candidate set further.
Figure~\ref{fig:preclustering} gives the algorithm.

Earlier frameworks further restricted which pairs could share
a cluster by comparing positive degrees and requiring many
common positive neighbors. They then modified an optimal
partition at small additional cost to respect those
restrictions~\cite{cohen-addad-23-preclustering,cao-24-lp}.
Our sampling analysis instead uses the total positive degree
of the free vertices. We therefore allow every compatible
cross-unit pair with a free endpoint. The construction
preserves every optimum, and normalization removes only
costs that all feasible unit-preserving partitions must pay.

\subsection{Atoms and 2-expansion}

For $Q\subseteq V$ and $v\in Q$, define
\[
 e_Q(v)=e^-(v,Q)+e^+(v,V\setminus Q),\qquad
 \delta_Q=\max_{v\in Q}e_Q(v).
\]
Thus $e_Q(v)$ counts the disagreements incident to $v$
when $Q$ is a cluster: negative pairs inside $Q$ and positive
pairs leaving $Q$. A \emph{dangerous outsider}
is a vertex $w\notin Q$ with $e^+(w,Q)\ge|Q|/3$ and a
cannot-link to some vertex of $Q$.

A \emph{$2$-expansion} from a set $W$ into $Q$ assigns two
distinct neighbors in $Q$ to every vertex of $W$, with no
assigned vertex reused. Here adjacency means a cannot-link.
Equivalently, duplicating every vertex of $W$ yields a
bipartite graph with a matching saturating all copies.

\twoexpansion*
\begin{proof}
Replace every vertex of $R$ by two copies with the same
neighbors, and find a maximum matching in the resulting
bipartite graph. Starting at all unmatched vertices of $L$,
follow alternating paths: unmatched edges from left to right
and matched edges from right to left. Let $Q$ be the reached
vertices of $L$, and let $Z$ consist of the vertices of $R$
with a reached copy.

Every neighbor of a reached left vertex is reached: an
unmatched edge can be followed, while its matched neighbor,
if any, is its predecessor on an alternating path.
The two copies of a right vertex have identical neighborhoods,
so either both are reached or neither is. Hence $N(Q)=Z$.
No reached right copy is unmatched, since otherwise an
augmenting path would exist. The matching edges incident to
the two copies of every vertex of $Z$ therefore give the
required $2$-expansion into $Q$.

Every vertex in $L\setminus Q$ is matched to a copy of a vertex
in $R\setminus Z$, yielding
$|L\setminus Q|\le2|R\setminus Z|$.
Isolated left vertices are unmatched and belong to $Q$.
The algorithm returns all vertices of $Q$, including those
not used as witnesses by $M$. All steps take polynomial time.
\end{proof}

We are now ready to state a sufficient condition for being an atom.  
The density condition is sufficient to prove preservation in
ordinary correlation clustering. The corresponding exchange
arguments were given in~\cite[Lemma~14]{cohen-addad-23-preclustering} and
\cite[Lemmas~15--16]{cao-24-lp}; the latter established
preservation by every optimum.
Here the additional certificate controls incompatible outsiders.
We will show that, if an optimum splits $Q$, every incompatible
outsider in the cluster containing its largest piece is
dangerous. Its two witnesses must lie in the other pieces of
$Q$, because that cluster is feasible.
Thus each expelled
outsider can be charged to two distinct vertices that we move into the cluster.

\begin{lemma}[Atom preservation]\label{lem:safecore}
Suppose $|Q|\ge2$, $Q$ has no internal cannot-link pair,
$\delta_Q<|Q|/20$, and the cannot-link graph admits a
$2$-expansion from the dangerous outsiders into $Q$.
Then $Q$ is an atom. In every optimal feasible partition,
its cluster contains at most $2\delta_Q$ vertices outside $Q$.
Moreover, any two disjoint sets satisfying these hypotheses
lie in different clusters in every optimum.
\end{lemma}
\begin{proof}
Suppose an optimal feasible partition splits $Q$.
Let $C$ contain its largest piece, breaking ties arbitrarily,
and put $S=Q\cap C$ and $R=Q\setminus C$.
Writing $Q_1,\ldots,Q_k$ for the intersections of $Q$ with
the optimal clusters, we have $|Q_i|\le|S|$, so the number
of separated pairs within $Q$ is at least
\[
 \frac{|Q|^2-\sum_i|Q_i|^2}{2}
 \ge\frac{|Q|(|Q|-|S|)}2
 =\frac{|Q||R|}{2}.
\]
The number of negative pairs among them is at most
$|R|\delta_Q$, since every such pair has an endpoint in $R$.
Making $Q$ a separate cluster is feasible and loses at most
$|Q|\delta_Q$ in external positive edges.
Optimality gives
\[
 0\le-\frac{|Q||R|}{2}
       +2|R|\delta_Q+|Q|\delta_Q.
\]
If $\delta_Q=0$, this already contradicts $R\ne\varnothing$.
Otherwise,
\[
 |R|\le\frac{2|Q|\delta_Q}{|Q|-4\delta_Q}<3\delta_Q,
 \qquad |S|>|Q|-3\delta_Q.
\]

Put $T=C\setminus Q$. Separating $S$ from $T$ is feasible
and changes cost by $2e^+(S,T)-|S||T|$. Since
$e^+(S,T)\le|S|\delta_Q$, optimality implies
$|T|\le2\delta_Q$. Isolating any $w\in T$ also cannot
improve the partition, hence
\[
 e^+(w,S)\ge\frac{|S|-|T|+1}{2}
 >\frac{|Q|-5\delta_Q}{2}>\frac{|Q|}{3}.
\]
Partition $T$ into outsiders $X$ incompatible with $Q$ and
outsiders $Y$ compatible with all of $Q$.
By the preceding inequality, every vertex of
$X$ is dangerous. Its two assigned cannot-link witnesses
lie in $R$, since $C$ is feasible. The $2$-expansion gives
\[
 |X|\le|R|/2,\qquad |Y|\le2\delta_Q.
\]

Replace $C$ by $Q\cup Y$ and $X$, removing the vertices
of $R$ from their former clusters and discarding empty clusters.
The new partition is feasible: $Q$ has no internal cannot-link,
$Y$ is compatible with $Q$, and both $X$ and $Y$ were subsets
of $C$.
The changes in cost from reuniting $Q$, removing $R$ from
its former clusters, merging $R$ with $Y$, and separating
$X$ from $S\cup Y$ are bounded above respectively by
\[
 -|R||S|+2|R|\delta_Q,\qquad |R|\delta_Q,
 \qquad |R||Y|,\qquad |X|(|S|+|Y|).
\]
Their sum is at most
\[
 \begin{aligned}
 \Delta
 &\le3|R|\delta_Q-(|R|-|X|)|S|
                     +(|R|+|X|)|Y|\\
 &\le6|R|\delta_Q-\frac{|R||S|}{2}<0,
 \end{aligned}
\]
because $|S|>|Q|-3\delta_Q>17\delta_Q$.
This contradicts optimality.

Once $Q$ is intact, separating it from the outsiders $T$
in its cluster yields
\[
 0\le2e^+(Q,T)-|Q||T|
 \le2|Q|\delta_Q-|Q||T|,
\]
so $|T|\le2\delta_Q$. If disjoint sets $Q,B$ satisfying
the hypotheses shared an optimal cluster, we would have both
\[
 |B|\le2\delta_Q<|Q|/10,
 \qquad
 |Q|\le2\delta_B<|B|/10,
\]
a contradiction.
\end{proof}

\subsection{Finding the atoms}

We now construct the atoms and bound the total positive degree of the free vertices.  Every compatible pair of distinct units has a free endpoint, so the number of positive pairs between compatible units is at most $\sum_{v\in F}e^+(v, V)$. Lemma~\ref{lem:weighted-error} uses this bound and the LP cost to control sampling error.

The degree bound includes all free vertices, including those removed from candidate sets in the additional trimming step for $2$-expansion.

\begin{lemma}\label{lem:atom-extraction}
Let $I$ be the instance entering preclustering.
One can construct disjoint atoms and a free-vertex set $F$
in polynomial time such that
\begin{equation}\label{eq:free-budget}
 \sum_{v\in F}e^+(v, V)\le C_{\mathrm{pre}}\OPT(I).
\end{equation}
\end{lemma}
\begin{proof}
Take the feasible partition $\calP$ supplied by the deterministic
version of Theorem~\ref{thm:metric-repulsion}, of cost
$\cost_I(\calP)\le3\OPT(I)$.
For each cluster $S\in\calP$, put
\[
 q_S=2e^-(S)+e^+(S,V\setminus S)
     =\sum_{v\in S}e_S(v),
 \qquad
 \sum_{S\in\calP}q_S=2\cost_I(\calP).
\]
If $|S|=1$, mark its vertex free; its positive degree is
exactly $q_S$. For the remaining clusters, if
$q_S\ge|S|^2/10000$, mark all vertices of $S$ free.
Otherwise, remove the vertices with $e_S(v)>|S|/100$,
mark them free, and let $S'$ be the remaining set. Then
\[
 |S\setminus S'|\le\frac{100q_S}{|S|}.
\]
Let
\[
 W_S=\{w\notin S:e^+(w,S)\ge|S|/4\}.
\]
Every vertex of $W_S$ contributes at least $|S|/4$ positive
edges to the boundary of $S$. Hence
\[
 \frac{|S|}{4}|W_S|
 \le\sum_{w\in W_S}e^+(w,S)
 \le e^+(S,V\setminus S)\le q_S,
 \qquad
 |W_S|\le\frac{4q_S}{|S|}.
\]
Form the bipartite cannot-link graph with parts $S',W_S$.
Lemma~\ref{lem:two-expansion} finds $Q\subseteq S'$ and
$Z\subseteq W_S$ with $N(Q)=Z$, a $2$-expansion
from $Z$ into $Q$, and
\[
 |S'\setminus Q|\le2|W_S\setminus Z|\le2|W_S|.
\]
Here $N(Q)$ is the neighborhood in this bipartite graph.
Consequently,
\[
 |S\setminus Q|\le\frac{108q_S}{|S|}<0.0108|S|,
 \qquad
 |Q|>0.9892|S|.
\]
Since $|S|\ge2$, the latter inequality also gives $|Q|\ge2$.
For every $v\in Q$,
\[
 e_Q(v)\le e_S(v)+|S\setminus Q|
 <0.0208|S|<|Q|/20.
\]

Since $Q\subseteq S$ and $S$ belongs to a feasible partition,
$Q$ contains no internal cannot-link, and every dangerous
outsider of $Q$ lies outside $S$.
For any such outsider $v$,
\[
 e^+(v,S)\ge e^+(v,Q)\ge|Q|/3>|S|/4.
\]
Thus, $v\in W_S$.
Since $v$ has a cannot-link neighbor in $Q$,
we have $v\in N(Q)=Z$.
Thus the $2$-expansion from $Z$ into $Q$
assigns two distinct, non-reused witnesses in $Q$ to every
dangerous outsider of~$Q$.  Lemma~\ref{lem:safecore} therefore applies.
Retain $Q$ as an atom and mark $S\setminus Q$ free.
The retained sets are disjoint because they come from distinct
clusters of $\calP$, and Lemma~\ref{lem:safecore} ensures that
every optimum places them in different clusters.

If all vertices of a cluster $S$ with $|S|\ge2$ are marked free,
their total positive degree is at most
\[
 |S|^2+e^+(S,V\setminus S)
 \le10000q_S+q_S=10001q_S.
\]
If $Q$ is retained, the free vertices in $S\setminus Q$
have sum of positive degrees at most
\[
 |S|\,|S\setminus Q|+e^+(S,V\setminus S)\le109q_S.
\]
Summing over the clusters of $\calP$ gives
\[
 \sum_{v\in F}e^+(v, V)
 \le10001\sum_{S\in\calP}q_S
 =20002\cost_I(\calP)
 \le6\cdot10001\OPT(I). \qedhere
\]
\end{proof}

As the final step of preclustering, first add cannot-links
between every pair of distinct atoms.
Lemma~\ref{lem:safecore} and the construction above show that
every optimal partition of $I$ already separates these atoms.
Next propagate every cannot-link between whole units:
if one pair between two units is a cannot-link, make all
pairs between them cannot-links.
Let $H$ denote this augmented and propagated set.

Apply the fixed-cost relabeling step of normalization
(Definition~\ref{def:normalization}), retaining all vertices and units.  Write $I'$ for the residual instance.

\preclusteringresult*
\begin{proof}
Lemma~\ref{lem:atom-extraction} constructs the atoms and proves
\[
 \sum_{v\in F}e^+(v, V)\le C_{\mathrm{pre}}\OPT(I),
\]
where $G$ is the positive graph before normalization.
By Lemma~\ref{lem:safecore}, every optimal partition of $I$
keeps each atom whole and separates distinct atoms.
It therefore respects the added inter-atom cannot-links.
For a unit-preserving partition, satisfying a cannot-link is
equivalent to separating its endpoint units, so propagation
preserves feasibility on these partitions.
Consequently, every optimum of $I$ remains feasible after
the constraint changes, and every feasible unit-preserving
partition of $I'$ is feasible for $I$.

The fixed-cost identity from normalization (Definition~\ref{def:normalization}) gives, for every such partition,
\[
 \cost_I(\calP)-\cost_{I'}(\calP)=|H\cap E^+|,
\]
where $E^+$ in this display denotes the labels before
normalization.

It remains to show that every optimum of $I'$ also preserves
the atoms. The internal labels of an atom $Q$ are unchanged,
and its external positive neighborhood only shrinks.
Thus $\delta_Q<|Q|/20$ still holds.
Every outsider incompatible with $Q$ now has
$e^+(w,Q)=0$, so there are no dangerous outsiders.
Lemma~\ref{lem:safecore}, with an empty $2$-expansion,
shows that every optimum of $I'$ preserves $Q$ and that
distinct atoms lie in different optimal clusters.
Applying the cost identity to optima of $I$ and $I'$ gives
\[
 \OPT(I)-\OPT(I')=|H\cap E^+|\ge0,
\]
again using the positive labels before normalization.
This proves the cost identity and property~(iii).

Finally, propagation makes $H$ a union of whole cross-products of distinct units, and there is no internal cannot-link. Distinct atoms are incompatible, so every compatible pair of distinct units has a free endpoint. This proves property~(i). Atom preservation and separation prove property~(ii). All construction steps take polynomial time.
\end{proof}

From this point onward, $E^+$ and $E^-$ refer to the labels
of the residual instance $I'$, unless stated otherwise.
The cross-unit pairs that may still be co-clustered are
\[
 \left\{uv\in\binV:B(u)\ne B(v),\ uv\notin H\right\}.
\]
Distinct atoms are incompatible, so every such pair has
a free endpoint.

\section[Bounded sub-cluster LP and sampling]{The bounded sub-cluster LP and candidate sampling}
\label{sec:lp}

We use the bounded sub-cluster LP of Cao et al.~\cite{cao-24-lp},
which already respects the atoms produced by preclustering.
We index its inclusion variables by units and impose the
separations supplied by our constraint-aware preclustering.
We omit their additional cluster-size gap.
We solve this LP
once and use its solution to sample candidate sets. Repulsion
partitions these candidates into feasible clusters in
Section~\ref{sec:repair}.

First consider an integral partition into $k$ clusters, with
$m_s$ clusters of size $s$. Choose $s$ with probability $m_s/k$,
then choose a pivot vertex $a$ uniformly among the $s m_s$
vertices in those clusters. Each cluster is selected with
probability $1/k$. Conditioning on the pivot makes the inclusion
value of every unit zero or one, so independent draws return
exactly the selected cluster.

For a fractional solution, units may have conditional inclusion
values strictly between zero and one. Independent draws ignore
their joint inclusion values: two units with individual values
$1/2$ and joint value zero would be selected together with
probability $1/4$. To control this discrepancy, we first select
a few additional units, called \emph{seeds}, and condition on
their inclusion. A selected seed has inclusion value one, while
other inclusion values may increase or decrease. The final
candidate contains the pivot unit and every selected seed.

Cao et al.~\cite[Section~4.2]{cao-24-lp} used Raghavendra--Tan
correlated rounding after selecting the size and pivot. Our LP
retains the local consistency constraints needed for that
procedure, but their relative-error analysis also uses a
cluster-size gap, stated in property~\cite[(A1)]{cao-24-lp}, and a bound on the
number of admissible pairs. Our construction does not provide
these additional properties, so that analysis does not apply
directly. We instead analyze the positive-conditioning sampler in Figure~\ref{fig:sample}.

Throughout this section, $n=|V|$ and $[n]=\{1,\ldots,n\}$.
The sets $E^+$ and $E^-$ are the positive and negative pairs
after the label changes of Theorem~\ref{thm:pre}.
Write $I$ for the instance before those changes; the remaining instance $I'$ has optimal value $\OPT(I')\le\OPT(I)$. The difference is the fixed cost removed by normalization.

\subsection{LP properties and conditioning}
\label{sec:lp-definition}

The LP is defined in Section~\ref{sec:outline-lp}, and the sampler is given in Figure~\ref{fig:sample}.   Here we establish the identities needed for sampling and prove feasibility and the objective bound.

Constraint~\eqref{lp:SA} gives $y_{s,\{B,D\}}\ge0$ and
$y_{s,\{B\}}-y_{s,\{B,D\}}\ge0$. Summing over $s$ and using
coverage~\eqref{lp:coverage}, we obtain
$0\le\sum_s y_{s,\{B,D\}}\le\sum_s y_{s,\{B\}}=1$.
Thus the values defined by~\eqref{lp:pair} satisfy
\begin{align}
 0\le x_{BD}&\le1 &&(B,D\in\calU),\label{lp:bounds}\\
 x_{BB}&=0 &&(B\in\calU).\label{lp:nothing}
\end{align}
For three distinct units $B,D,E$, inclusion--exclusion gives
\[
 y_{s,\{D\}}-y_{s,\{B,D\}}-y_{s,\{D,E\}}
                  +y_{s,\{B,D,E\}}\ge0.
\]
Monotonicity gives $y_{s,\{B,D,E\}}\le y_{s,\{B,E\}}$.
Summing over $s$ and using coverage yields
$(1-x_{BD})+(1-x_{DE})\le1+(1-x_{BE})$, hence
\begin{equation}\label{lp:triangle}
 x_{BE}\le x_{BD}+x_{DE}\qquad(B,D,E\in\calU).
\end{equation}
Repeated-unit cases follow from the zero diagonal and
nonnegativity. Finally, incompatible units have distance one
by~\eqref{lp:support}. This proves that $x$ is indeed a feasible
unit-preserving metric. The proof uses variables indexed by
at most three units, explaining the condition $r\ge3$.

\begin{lemma} \label{lem:LP}
The unit-preserving LP of Definition~\ref{def:moment-lp}
is feasible and can be solved as a rational LP of size
$n^{O(r)}$.  The optimal LP objective value $L$ satisfies $L\le\OPT(I')$.
Furthermore, for every $s$ and every $\calT$ with
$|\calT|\le r-1$,
\begin{equation}\label{eq:size-obstruction}
 \left(s-\sum_{B\in \calT}|B|\right)y_{s, \calT}
 =\sum_{B\in\calU\setminus \calT}|B|y_{s, \calT\cup\{B\}}\ge0.
\end{equation}
In particular $y_{s, \calT}=0$ when
$\sum_{B\in \calT}|B|>s$.
\end{lemma}
\begin{proof}
Take an optimal partition $\calP^*$ of the remaining instance.
By Theorem~\ref{thm:pre}, it preserves units and respects all propagated cannot-links. For each $s,\calT$ let $y_{s, \calT}$ be the
\emph{number} of $s$-clusters containing every unit of $\calT$;
in particular, $y_{s, \varnothing}$ counts the $s$-clusters.
For disjoint $P,N$, inclusion--exclusion shows that the left side
of \eqref{lp:SA} is the number of $s$-clusters containing $P$
and avoiding $N$, so it is nonnegative.
For any $s$-cluster containing $\calT$, the sum of the sizes of
all its member units is exactly $s$, establishing
\eqref{lp:size}.  Coverage and pair linking hold because every
unit belongs to precisely one integral cluster.  The derived values $x_{BD}$ are exactly its separation indicators. Constraint~\eqref{lp:support} holds because $\calP^*$ is feasible.  
The optimal LP objective is no larger, proving the upper bound.

To obtain \eqref{eq:size-obstruction}, separate the terms
$B\in \calT$ from \eqref{lp:size}; for these,
$y_{s, \calT\cup\{B\}}=y_{s, \calT}$.  The remaining terms are nonnegative
by \eqref{lp:SA}.  Finally the number of pairs $(P,N)$ with
$|P|+|N|\le r$ is $|\calU|^{O(r)}$, and there are $n$ size layers.
All coefficients have polynomial input bit length (the largest
coefficient $|B|$ is at most $n$).  Standard exact rational linear
programming therefore yields a polynomial-bit solution in time
$n^{O(r)}$.
\end{proof}

For sampling, assume $r\ge6$ is an even integer and fix a rational optimal solution $y$ and its derived metric $x$. Its level is chosen in Section~\ref{sec:completion}.
Abbreviate
\begin{equation}\label{eq:Lplusminus}
 L_+=\sum_{uv\in E^+}x_{uv},\qquad
 L_-=\sum_{uv\in E^-}(1-x_{uv}),\qquad L=L_++L_-.
\end{equation}
Also write $y_\calT=\sum_{s=1}^n y_{s, \calT}$.

Recall that $y_{s, \varnothing}$ is the fractional number of $s$-clusters. Put
\begin{equation}\label{eq:cluster-count}
 Z=\sum_{s=1}^n y_{s, \varnothing}.
\end{equation}
Thus $Z$ is the fractional number of clusters represented by the
LP solution; for an integral partition into $k$ clusters, $Z=k$.
Summing the size identities~\eqref{lp:size} and using coverage
\eqref{lp:coverage} gives
\begin{equation}\label{eq:cluster-count-basic}
  1\le Z=\sum_{s=1}^n y_{s, \varnothing} \le \sum_{s=1}^n s y_{s, \varnothing}=n.
\end{equation}

The probability of selecting size $s$ and pivot $a$ is
\begin{equation}\label{eq:pivot-law}
  \lambda_{s,a} =
\frac{y_{s, \varnothing}}Z \frac{y_{s, \{B(a)\}}}{s y_{s, \varnothing}}
  = \frac{y_{s, \{B(a)\}}}{s Z}.
\end{equation}
The size identities show that both probability distributions in
Steps~1 and~2 sum to one. Formula~\eqref{eq:pivot-law} extends
the uniform cluster selection described above to fractional solutions.

Fix a sampled size $s$ and pivot $a$. As in Step~4, let
$\calW$ be the family of units other than $B(a)$ that are
compatible with $B(a)$, and put
\[
 t=s-|B(a)|,\qquad W=\bigcup_{D\in\calW}D.
\]
The values initialized in Step~5 describe inclusion conditional
on the pivot unit: $z_{\{D\}}$ is the inclusion value of $D$,
and $z_{\{D,E\}}$ is the joint inclusion value of $D,E$.
Each seed update conditions these values on the selected unit.
Selecting an already chosen seed leaves every retained entry
unchanged, but still counts as an update. All entries in
Step~7.2 are computed from the old table; Step~7.3 replaces them
only after those computations are complete. After $j$ updates,
the stored entries have index sets of size at most $r-1-j$.

Conditioning on a seed $D$ sets the inclusion value of every
unit $E$ with $x_{DE}=1$ to zero, because $z_{\{D,E\}}=0$.
Later updates preserve this exclusion. The final independent
draws can nevertheless select two incompatible non-seed units,
which Repulsion will separate.

For the analysis, write $z_D=z_{\{D\}}$ and
$z_{DE}=z_{\{D,E\}}$. For vertices $i,j\in W$, write
$z_i=z_{B(i)}$ and $z_{ij}=z_{\{B(i),B(j)\}}$.
Vertices in the same unit share a selection, so $z_{ij}=z_i$
when $B(i)=B(j)$, including $i=j$. Extend this notation by
setting the inclusion of the pivot unit to one and the inclusion
of vertices outside $B(a)\cup W$ to zero. These values describe
consistent probabilities on small sets; a distribution realizing
all of them simultaneously is not assumed.

An integral $s$-cluster containing the pivot has exactly $t$
vertices outside its unit. The following invariants express
this count through conditional LP values, including after seed updates. Taking $\calT=\varnothing$ in \eqref{eq:fixed-t} normalizes the seed probabilities and shows that the final candidate has conditional expected size $s$, although its actual size may differ. For analysis, the conditional formulas below
also define entries not stored by the algorithm.

\begin{enumerate}[label=(\roman*)]
\item \textbf{Initial conditioning.}
  Initially,
  \[
    z_{\calT}
    =\frac{y_{s,\calT\cup\{B(a)\}}}{y_{s,\{B(a)\}}}
    \qquad
    (\calT\subseteq\calU,\ |\calT\cup\{B(a)\}|\le r).
  \]
  These values satisfy $z_{\varnothing}=z_{\{B(a)\}}=1$ and
  $z_{\calT}=0$ if $\calT\not\subseteq\calW\cup\{B(a)\}$.

\item \textbf{Invariance under seed updates.}
  After $j\ge0$ seed updates, let $\calS$ consist of the pivot
  unit and the distinct seeds selected so far.
  Then $y_{s,\calS}>0$, and
  \[
    z_{\calT}=\frac{y_{s,\calS\cup\calT}}{y_{s,\calS}}
    \qquad(\calT\subseteq\calW,\ |\calS\cup\calT|\le r).
  \]
  For $\calT\subseteq\calW$ with $|\calS\cup\calT|\le r-1$,
  \begin{equation}\label{eq:fixed-t}
    \sum_{D\in\calW}|D|z_{\calT\cup\{D\}}=t z_{\calT}.
  \end{equation}
\end{enumerate}
\begin{lemma}\label{lem:conditioning-structure}
For every $s,a$ with $y_{s,\{B(a)\}}>0$, we have $t\ge0$.
The initialization and seed updates in Figure~\ref{fig:sample}
satisfy the conditional formulas above, with positive denominators.
Identity~\eqref{eq:fixed-t} holds whenever
$|\calS\cup\calT|\le r-1$.
\end{lemma}
\begin{proof}
  (i) Initial conditioning.
By \eqref{eq:size-obstruction} and $y_{s,\{B(a)\}}>0$,
we have $s\ge|B(a)|$, so $t\ge0$.
The initial ratios agree with Step~5 and give
$z_{\varnothing}=z_{\{B(a)\}}=1$.

If $D\notin\calW\cup\{B(a)\}$, then $D$ is incompatible
with the pivot unit. Constraint~\eqref{lp:support} gives
$y_{s,\{B(a),D\}}=0$.
Nonnegativity and monotonicity from \eqref{lp:SA} imply
that every variable whose index set contains both units
also vanishes. This proves the asserted zero inclusion values.

\smallskip
\noindent  (ii) Invariance under seed updates.
Initially $\calS=\{B(a)\}$, so the conditional formula holds
and $y_{s,\calS}>0$.
Suppose these properties hold before an update.
A selected seed $D$ has $z_{\{D\}}>0$, and hence
\[
 y_{s,\calS\cup\{D\}}
 =y_{s,\calS}z_{\{D\}}>0.
\]
The update gives
\[
 z_{\calT}^{\mathrm{new}}
 =\frac{z_{\calT\cup\{D\}}}{z_{\{D\}}}
 =\frac{y_{s,\calS\cup\{D\}\cup\calT}}
        {y_{s,\calS\cup\{D\}}},
\]
which is the required formula after adding $D$ to $\calS$.
If $D$ was selected earlier, then $D\in\calS$ and the
values remain unchanged.

It remains to prove \eqref{eq:fixed-t}.
For $\calT\subseteq\calW$ with $|\calS\cup\calT|\le r-1$,
apply \eqref{lp:size} to $\calS\cup\calT$.
The pivot unit belongs to $\calS$, and terms involving
units incompatible with it vanish by the argument above.
Thus
\[
 s\,y_{s,\calS\cup\calT}
 =|B(a)|\,y_{s,\calS\cup\calT}
  +\sum_{D\in\calW}|D|\,
       y_{s,\calS\cup\calT\cup\{D\}}.
\]
Subtracting the pivot term and dividing by $y_{s,\calS}>0$
proves \eqref{eq:fixed-t}, including the initial case $j=0$.
\end{proof}

For every pair $u,v$, define the LP target
\begin{equation}\label{eq:ideal-pairs}
 q^0_{uv}=\frac{1-x_{uv}}{Z}.
\end{equation}
This number depends only on the fixed LP solution.  For an integral partition, it is the probability that
both vertices belong to a uniformly selected cluster.
Separately, let
\[
 q_{uv}=\Prob[u,v\in C],
 \qquad C\text{ sampled by Figure~\ref{fig:sample}}.
\]
This is the actual joint inclusion probability of the sampler.
In the next lemma, $z_v$ and $z_{uv}$ denote the final conditional
values before Step~8. If Step~3 returns $B(a)$ instead, set
$z_v=\ind[v\in B(a)]$ and $z_{uv}=\ind[u,v\in B(a)]$.

\begin{lemma} \label{lem:marginals}
  The sampler in Figure~\ref{fig:sample} queries variables indexed by at most $r$ units.
  In each seed step $j$, every updated entry preserves its conditional expectation:
  \begin{equation}\label{eq:martingale}
    \E[z_\calT^{\mathrm{new}} \mid \text{current table}] = z_\calT
    \qquad \text{for all } \calT \subseteq \calW \text{ with } |\calT| \le r - 1 - j.
  \end{equation}
  Averaged over all random choices of the sampler, the candidate $C$ satisfies
  \begin{equation}\label{eq:singleton}
    \Prob[v \in C] = \Prob[B(v) \subseteq C] = \frac{1}{Z} \qquad (v \in V).
  \end{equation}
The final conditional inclusion values immediately before
Step~8 satisfy
\[
 \E[z_v]=\frac1Z,\qquad \E[z_{uv}]=q^0_{uv}.
\]
If $B(u)=B(v)$, then $\Prob[u,v\in C]=q^0_{uv}=1/Z$.
\end{lemma}
\begin{proof}
Fix a positive-probability choice $(s,a)$.
By the seed-update invariance in
Lemma~\ref{lem:conditioning-structure}, the current values
satisfy, after every seed history,
\[
 \sum_{D\in\calW}|D|z_D=t,
 \qquad
 \sum_{D\in\calW}|D|z_{\calT\cup\{D\}}=t z_{\calT}.
\]
Here and below sums over $D\in \calW$ range over eligible units. Local assignment nonnegativity gives
$0\le z_D\le1$ and $0\le z_{\calT\cup\{D\}}\le z_D$.
Consequently the seed probabilities sum to one, the denominator
for every selected seed is positive, and its conditioning produces
valid local probabilities.  For $t > 0$,
\[
 \E[z_\calT^{\mathrm{new}}\mid\text{current table}]
 =\sum_{D:\,z_D>0}\frac{|D|z_D}{t}
                         \frac{z_{\calT\cup\{D\}}}{z_D}
 =z_\calT.
\]
Terms with $z_D=0$ have zero joint inclusion value and are omitted.
If $t=0$, the size identity instead forces all singleton inclusion probabilities
in $\calW$ to vanish, justifying the early return.

For the estimates below, extend the seed sequence to $r/2$ steps
for analysis and use its first $K$ steps for the actual output.
The pivot, at most $r/2-1$ seeds, two queried units, and one
additional unit for a size identity involve at most $r/2+3\le r$
coordinates. The extra step used in the variance calculation
also fits within the available order. Thus all values used below
are supplied by the LP.

By \eqref{eq:martingale}, the expected final singleton inclusion probability
given $(s,a)$ equals its value immediately after conditioning on
the pivot. Independent sampling preserves that singleton value.
Using \eqref{eq:pivot-law}, the size identities, and coverage,
we obtain
\[
 \Prob[v\in C]
 =\sum_{s,a}\lambda_{s,a}
       \frac{y_{s, \{B(a),B(v)\}}}{y_{s, \{B(a)\}}}
 =\frac1Z\sum_s\frac1s\sum_{a\in V}
                         y_{s, \{B(a),B(v)\}}
 =\frac1Z\sum_s y_{s, \{B(v)\}}=1/Z.
\]
All sums over $(s,a)$ omit zero-probability choices. For the
joint LP value, \eqref{eq:martingale} likewise preserves its
expectation during the seed steps. Thus
\[
 \E[z_{uv}]
 =\sum_{s,a}\lambda_{s,a}
       \frac{y_{s, \{B(a),B(u),B(v)\}}}{y_{s, \{B(a)\}}}
 =\frac1Z\sum_s y_{s, \{B(u),B(v)\}}
 =\frac{1-x_{uv}}{Z}=q^0_{uv}.
\]
This calculation proves that the averaged conditional value equals
the previously defined LP target; it does not define that target.
Every unit is included or excluded as a whole. For $v\in B$,
the events $v\in C$ and $B\subseteq C$ are therefore identical,
which proves the unit inclusion formula. It also proves that
vertices in the same unit have joint inclusion probability $1/Z$.
\end{proof}

\subsection{Sampling error}
\label{sec:sampling-error}
Step~8 does not change the table. Given the earlier choices, it
selects distinct units independently using their final individual
probabilities. For vertices in different units, therefore,
\[
 q_{uv}=\E[z_u z_v],\qquad q^0_{uv}=\E[z_{uv}].
\]
Although $\E[z_u]=\E[z_v]=1/Z$, the random values $z_u,z_v$
depend on the same size, pivot, and seeds, so in general
$\E[z_u z_v]\ne1/Z^2$. The error comes from using the conditional
product $z_u z_v$ rather than the LP joint value $z_{uv}$.
For $u,v$ with $B(u)\ne B(v)$, define
\[
  d_{uv}
  = \frac{\E[z_uz_v]-\E[z_{uv}]}{\E[z_u]}
 =Z(q_{uv}-q^0_{uv})
\]
as the normalized difference between the actual pair probability and its LP target.
Pairs within a unit have exact joint inclusion probability
$q_{uv}=q^0_{uv}=1/Z$ and contribute no sampling error.
Let
\[
  Y=\sum_{\substack{u<v\\B(u)\ne B(v)}}(1-x_{uv})
\]
denote the total LP co-clustering mass on pairs in different
units. For an integral partition, it counts such pairs whose
endpoints share a cluster, regardless of sign.
Lemma~\ref{lem:weighted-error} bounds $Y$ by the free-vertex
degree budget plus the LP cost, and hence by a constant multiple
of $\OPT(I)$.

The next lemma bounds both squared errors and their signed sum.
Its proof sums the expected increases of two bounded quantities:
the squared norm of inclusion values for squared error, and the
total size of distinct seeds for signed error.
These sums range over unordered pairs in different units,
including pairs in $H$. The squared bound
will control error weighted by positive separation costs through
Cauchy--Schwarz. The signed bound controls the remaining terms
without requiring a small absolute error summed over all pairs.

Let $\mathscr D$ denote the distribution of the candidate $C$ returned by Figure~\ref{fig:sample} for the fixed LP solution $y$, averaging over all random choices in the procedure.
Let
\[
 D_s=\sum_{\substack{uv\notin H\\B(u)\ne B(v)}}d_{uv}
\]
be the signed error over non-hard cross-unit pairs. This sum
need not be nonnegative.

\begin{lemma}[Sampling discrepancy bounds]\label{lem:discrepancy}
The distribution $\mathscr D$ satisfies
\begin{align}
 \sum_{\substack{u<v\\B(u)\ne B(v)}}d_{uv}^2
 &\le\frac{2Y}{r},\label{eq:D2}\\
 0\le\sum_{\substack{u<v\\B(u)\ne B(v)}}d_{uv}
 &\le\frac{2Y}{r},\qquad D_s\le\frac{2Y}{r}.
 \label{eq:signed}
\end{align}
\end{lemma}
\begin{proof}
Fix a size and pivot $(s,a)$ with $t=s-|B(a)|>0$.
Extend the seed sequence to $r/2$ updates for analysis,
independently of the uniform stopping index $K$, and let
$\mathcal H_k$ record the first $k$ seeds. Choices with $t=0$
have no sampling error and contribute zero to the averages below.

\emph{Squared errors.}
Write $z^{(k)}=(z_i^{(k)})_{i\in W}$. At a fixed history,
suppress the superscript and put
\[
 c_{ij}=z_{ij}-z_i z_j,\qquad
 \mathcal E_k=\sum_{i,j\in W}\frac{c_{ij}^2}{z_j},
\]
where a summand with $z_j=0$ is zero; local nonnegativity
then also gives $z_{ij}=0$. Selecting a seed is equivalent
to selecting a vertex $j\in W$ with probability $z_j/t$.
Its update changes $z_i$ by $c_{ij}/z_j$, with conditional
mean zero by~\eqref{eq:martingale}. Consequently,
\begin{equation}\label{eq:norm-growth}
 \E\!\left[\|z^{(k+1)}\|_2^2-\|z^{(k)}\|_2^2
               \,\middle|\,\mathcal H_k\right]
 =\frac{\mathcal E_k}{t}.
\end{equation}
Since $0\le z_i\le1$ and $\sum_{i\in W}z_i=t$,
we have $0\le\|z^{(k)}\|_2^2\le t$. Summing the expected
increases and averaging over $K$ yields
\begin{equation}\label{eq:layer-variance}
 \E[\mathcal E_K\mid s,a]
 =\frac{2t}{r}\left(
   \E[\|z^{(r/2)}\|_2^2\mid s,a]-\|z^{(0)}\|_2^2\right)
 \le\frac{2t^2}{r}.
\end{equation}
To average over sizes and pivots, the size identities give
\begin{equation}\label{eq:Y-identity}
 \sum_{s,a}y_{s,\{B(a)\}}(s-|B(a)|)
 =\sum_{\substack{(a,j)\\B(a)\ne B(j)}}
      y_{\{B(a),B(j)\}}=2Y,
\end{equation}
where the middle sum is over ordered vertex pairs. Indeed,
apply~\eqref{lp:size} to $\{B(a)\}$ and subtract the pivot
unit's contribution. Using $\lambda_{s,a}=y_{s,\{B(a)\}}/(sZ)$
and $t\le s$, we obtain
\begin{equation}\label{eq:Abar}
 \overline{\mathcal E}
 =\sum_{s,a}\lambda_{s,a}\E[\mathcal E_K\mid s,a]
 \le\frac{2}{rZ}\sum_{s,a}y_{s,\{B(a)\}}t
 =\frac{4Y}{rZ}.
\end{equation}
For the final conditional values, set $c_{uv}=0$ unless
$u,v\in W$; outside this case one endpoint has deterministic
inclusion and the conditional sampling error is zero.
For a cross-unit pair, therefore,
$q_{uv}-q^0_{uv}=-\E[c_{uv}]$.
Cauchy--Schwarz and $\E[z_v]=1/Z$ from Lemma~\ref{lem:marginals} give
\[
 d_{uv}^2=Z^2(\E[c_{uv}])^2
 \le Z\E\!\left[\frac{c_{uv}^2}{z_v}\right].
\]
These expectations include the size, pivot, and seed choices.
Averaging the inequalities for the two orientations and summing
unordered cross-unit pairs counts a subset of the ordered terms
in $\overline{\mathcal E}$. Thus
\[
 \sum_{\substack{u<v\\B(u)\ne B(v)}}d_{uv}^2
 \le\frac{Z\overline{\mathcal E}}2\le\frac{2Y}{r}.
\]

\emph{Signed errors.}
Let $M_k$ be the total number of vertices in the distinct seed
units selected in the first $k$ updates. Selected seeds have
inclusion value one, so $M_0=0$ and $M_k\le t$.
If sampling stopped after $k$ updates, the independent unit
choices would have conditional size variance
\[
 V_k=\sum_{B\in\calW}|B|^2z_B^{(k)}(1-z_B^{(k)}).
\]
Selecting a previously unselected unit $B$ increases $M_k$
by $|B|$, whereas repeated selections leave it unchanged.
Already selected units have zero variance, and hence
\[
 t\E[M_{k+1}-M_k\mid\mathcal H_k]
 =\sum_{\substack{B\in\calW\\B\text{ not yet selected}}}
      |B|^2z_B^{(k)}\ge V_k.
\]
Summing the expected increases and averaging over $K$ gives
\begin{equation}\label{eq:conditional-size-variance}
 \E[V_K\mid s,a]
 \le\frac{2t}{r}\E[M_{r/2}-M_0\mid s,a]
 \le\frac{2t^2}{r}.
\end{equation}
Including the pivot coordinates, the conditional size identities
give $\sum_i z_i=s$ and $\sum_j z_{ij}=s z_i$.
Summing the latter over $i$ and using $z_{ii}=z_i$ yields
$\sum_{i<j}z_{ij}=\binom{s}{2}$.
The actual candidate has conditional mean size $s$ and variance
$V_K$, so
\[
 \E\!\left[\binom{|C|}{2}\,\middle|\,s,a,\mathcal H_K\right]
 -\sum_{i<j}z_{ij}=\frac{V_K}{2}.
\]
Averaging this identity and using the inclusion identities from Lemma~\ref{lem:marginals} gives
\begin{align*}
 0\le\sum_{\substack{u<v\\B(u)\ne B(v)}}d_{uv}
 &=\frac Z2\sum_{s,a}\lambda_{s,a}\E[V_K\mid s,a]\\
 &\le\frac1r\sum_{s,a}y_{s,\{B(a)\}}\frac{t^2}{s}
 \le\frac{2Y}{r},
\end{align*}
where the last step uses $t\le s$ and~\eqref{eq:Y-identity}.
For $uv\in H$, $x_{uv}=1$ gives $d_{uv}=Zq_{uv}\ge0$.
Removing these terms cannot increase the signed sum, proving
the bound on $D_s$.
\end{proof}

For the error analysis, define
\[
 S_F=\sum_{v\in F}e^+(v, V).
\]
Here $e^+$ is evaluated using the residual labels. Normalization
only removes positive edges, so Theorem~\ref{thm:pre} also gives
$S_F\le C_{\mathrm{pre}}\OPT(I)$ for this sum.

The Repulsion cost analysis also needs the total absolute error
on positive pairs. The next lemma obtains this bound by
Cauchy--Schwarz and makes the degree-charging argument above
formal: both $Y$ and the positive cross-unit pair count are
bounded by a constant multiple of $\OPT(I)$. The resulting
absolute error is $O(\OPT(I)/\sqrt r)$.
Define
\begin{equation}\label{eq:P2}
 P_+=|\{uv\in E^+:B(u)\ne B(v)\}|,
 \qquad D_+=\sum_{\substack{uv\in E^+\\B(u)\ne B(v)}}|d_{uv}|.
\end{equation}
Here $P_+$ counts positive pairs joining different units, and
$D_+$ is the total absolute normalized sampling error on positive
pairs. Pairs within a unit have exact joint inclusion probabilities and do not enter $D_+$.

\begin{lemma}[Error on positive pairs]
\label{lem:weighted-error}
The quantities defined above satisfy
\begin{equation}\label{eq:weighted-and-opt}
 D_+\le\sqrt{\frac{2YP_+}{r}},\qquad
 Y\le S_F+L_-,\qquad P_+\le S_F+L_+,
 \qquad Y,P_+\le(C_{\mathrm{pre}}+1)\OPT(I).
\end{equation}
\end{lemma}

\begin{proof}
No positive pair is hard, and $D_+$ includes only cross-unit pairs. Cauchy--Schwarz and Lemma~\ref{lem:discrepancy}
therefore give
\[
 D_+\le\sqrt{P_+\sum_{\substack{u<v\\B(u)\ne B(v)}}d_{uv}^2}
 \le\sqrt{\frac{2YP_+}{r}}.
\]
It remains to bound the two factors. Every positive cross-unit
pair is compatible, because normalization relabeled all positive
cannot-link pairs as negative. Such a pair has a free endpoint,
since distinct atoms are incompatible. Therefore $P_+\le S_F$.
Splitting the sum defining $Y$ by sign gives
\[
 \begin{aligned}
 Y&=\sum_{\substack{uv\in E^+\\B(u)\ne B(v)}}(1-x_{uv})
    +\sum_{\substack{uv\in E^-\\B(u)\ne B(v)}}(1-x_{uv})\\
  &\le P_++L_-\le S_F+L_-.
 \end{aligned}
\]
The stronger bound $P_+\le S_F$ also implies $P_+\le S_F+L_+$.
Finally, the bound $S_F\le C_{\mathrm{pre}}\OPT(I)$ established
above, optimality of the LP, and Lemma~\ref{lem:LP} give
$L_+,L_-\le L\le \OPT(I)$. This proves the stated bounds.
\end{proof}

\section[Repulsion and the expected cost]{Repulsion and the expected clustering cost}
\label{sec:repair}

Radius-one Repulsion partitions each sampled candidate into
feasible clusters. We first bound the cost of this operation,
then analyze the partition constructed by repeated sampling.
Finally, we bound the number of rounds and prove
Theorem~\ref{thm:main}.

\subsection{Partitioning a candidate by Repulsion}

All pair labels and costs below refer to the remaining instance
$I'$. For the common formulas for the two signs, let
$a_{uv}=\ind[uv\in E^+]$ be the binary adjacency indicator
of its positive graph.

Define the following function for the Repulsion cost bound:
\begin{equation}\label{eq:b-def}
 b(x)=\begin{cases}
     \dfrac{2x^2}{1-x},&0\le x<\dfrac{\sqrt5-1}{2},\\[6pt]
     2,&\dfrac{\sqrt5-1}{2}\le x\le1.
 \end{cases}
\end{equation}
By definition, $0\le b(x)\le2$. Two further properties serve
different parts of the analysis. First,
\begin{equation}\label{eq:b-quadratic}
 (1-x)b(x)\le2x^2\qquad(0\le x\le1).
\end{equation}
There is equality before the cap at two; after the cap, the
inequality follows from $x^2+x-1\ge0$ whenever $x\ge \dfrac{\sqrt5-1}{2}$.
This property will bound
the expected charge on a positive pair by twice its LP cost.
Second,
\begin{equation}\label{eq:b-complementary}
 b(s)+b(t)\ge2\qquad(s,t\in[0,1],\ s+t\ge1).
\end{equation}
This property makes the contributions of induced three-vertex
paths nonpositive in the Repulsion proof. It is immediate if
$b(s)=2$ or $b(t)=2$. Otherwise, Cauchy--Schwarz gives
\[
 \frac{s^2}{1-s}+\frac{t^2}{1-t}
 \ge\frac{(s+t)^2}{2-s-t}\ge1,
\]
and multiplying by two proves the claim.

We now partition a candidate using strict metric balls of radius one.
The center of a ball is separate from the pivot used to sample the
candidate. Define a nonnegative pair potential
\begin{equation}\label{eq:repair-potential}
 \phi_{uv}=\begin{cases}
 (1-a_{uv})+a_{uv}b(x_{uv}),&x_{uv}<1,\\
 a_{uv},&x_{uv}=1,
 \end{cases}
 \qquad \Phi(S)=\sum_{uv\in\binom S2}\phi_{uv}.
\end{equation}
For a center $v\in S$, put $A_v=\{u\in S:x_{uv}<1\}$. The
boundary cost of separating $A_v$ from the rest, plus the two new
potentials minus the old potential, is
\begin{align}\label{eq:repair-net}
 \Delta_v
 &=e^+(A_v,S\setminus A_v)+\Phi(A_v)+\Phi(S\setminus A_v)-\Phi(S)
 \nonumber\\
 &=\sum_{\substack{u\in A_v,\,w\in S\setminus A_v\\x_{uw}<1}}
                  [a_{uw}(2-b(x_{uw}))-1].
\end{align}
Pairs of distance one contribute zero because their potential
already equals their separation cost.

Apply the following recursion, initially with $S=D$.
\begin{enumerate}[label=\arabic*.,leftmargin=*]
\item If $S$ is empty, return no clusters. If every pair in $S$
      has distance less than one, output $S$ as one cluster.
\item Otherwise, compute $\Delta_v$ for every center $v\in S$
      with $A_v\ne S$, and choose one minimizing $\Delta_v$.
      Break ties deterministically.
\item Separate $A_v$ from $S\setminus A_v$ and recurse on both parts.
\end{enumerate}

The next lemma shows that the recursion is feasible and that its
cost is paid by the initial potential. A split may cut positive
pairs, but an appropriate center makes the decrease in potential
cover their cost. The surviving negative pairs in terminal clusters
are then paid for by the remaining potential. This is a bound on
the total cost, rather than a separate bound on each pair's cost.
\begin{lemma}[Repulsion bound]\label{lem:repair}
For every union of units $D$, Repulsion partitions $D$ in polynomial time, preserves each unit, and separates every cannot-link pair.
Its total internal cost in the remaining instance is at most
\begin{equation}\label{eq:repair-charge}
 \Phi(D)\le\sum_{uv\in\binom D2\setminus H}
                  [(1-a_{uv})+a_{uv}b(x_{uv})].
\end{equation}
\end{lemma}
\begin{proof}
Form the graph on $S$ whose edges are the distinct pairs of distance
less than one. Its closed neighborhoods are the balls $A_v$.
Sum $\Delta_v$ over all centers, allowing $A_v=S$, whose contribution
is zero. Each summand in \eqref{eq:repair-net} belongs to an induced
path of three vertices: the center is one endpoint, the short pair
being cut is the opposite edge, and the missing pair has distance
one. Thus an induced path $u-v-w$ contributes
\[
 a_{uv}(2-b(x_{uv}))-1+a_{vw}(2-b(x_{vw}))-1.
\]
The triangle inequality gives $x_{uv}+x_{vw}\ge x_{uw}=1$.
Since $2-b\ge0$ and $a_{uv},a_{vw}\le1$, this contribution is at most
\[
 2-b(x_{uv})-b(x_{vw})\le0
\]
by \eqref{eq:b-complementary}. Therefore $\sum_v\Delta_v\le0$.
If the graph is not complete, a proper ball exists; balls equal
to $S$ contribute zero. Some proper ball consequently has
$\Delta_v\le0$, and the minimizing center chosen by the algorithm
has this property.

At each split, the boundary cost plus the two new potentials is
at most the parent potential. In a final block every pair has
$x_{uv}<1$, and its co-clustering cost $1-a_{uv}$ is at most
$\phi_{uv}$. Induction over the recursion proves that the total
cost is at most $\Phi(D)$. Final blocks contain no pair of distance
one and hence no hard pair. If $u,w$ belong to the same unit,
$x_{uw}=0$ implies $x_{vu}=x_{vw}$ for every center $v$, so all
balls, and all recursive blocks, are unions of units.

For a hard pair, $x_{uv}=1$ and $a_{uv}=0$, so its potential
is zero. A compatible distance-one pair has potential $a_{uv}$,
which is at most $1-a_{uv}+a_{uv}b(1)=1+a_{uv}$.
All other pairs have exactly the charge displayed in
\eqref{eq:repair-charge}. This proves that inequality.
There are at most $|D|-1$ proper splits, each found by evaluating
at most $|D|$ centers, so the procedure is polynomial-time.
\end{proof}

\subsection{Constructing and analyzing the partition}
After the single LP solve, independently sample $C_1,C_2,\ldots$
from the \emph{same} distribution $\mathscr D$ defined in
Section~\ref{sec:sampling-error}.  Maintain a set $S$ of unprocessed
vertices.  In each round first put $D=C_i\cap S$; if $D$ is
nonempty, separate it from $S\setminus D$, apply Repulsion to \emph{this $D$}
using Lemma~\ref{lem:repair}, output its clusters and
remove all vertices of $D$.  This order (intersection before repair)
is essential to the pairwise first-hit calculation.  The
distribution and LP solution are never recomputed during rounding.
Moreover, every pair with LP distance one is separated in the
output: either its endpoints first appear in different rounds,
or Repulsion separates them in the round where they first appear
together. This includes distinct atoms and all other
separations prescribed by the LP.

To analyze these rounds, follow a pair until the first candidate
containing at least one endpoint. If it contains only one, the
pair is separated immediately. If it contains both, the pair
enters the potential bound for Repulsion in that round. The next
lemma combines these two cases into the function $F_a(x,y)$,
where $x$ is the LP distance and $y=Zq_{uv}$ is the scaled joint
inclusion probability. Its second inequality separates the ideal
charge at $y=1-x$ from the sampling error $d=y-(1-x)$.

\begin{lemma}[Cost at the first round meeting a pair]
\label{lem:firsthit}
Let $\mathrm{ALG}_\infty$ denote the partition obtained by
continuing these rounds indefinitely.
It terminates with probability $1$ (every vertex is eventually processed), producing a feasible partition of expected cost bounded by
\begin{equation}\label{eq:firsthit-intermediate}
 \E\cost_{I'}(\mathrm{ALG}_\infty)
 \le\sum_{uv\notin H} F_{a_{uv}}(x_{uv},Zq_{uv}),
\end{equation}
where
\begin{equation}\label{eq:F-def}
 F_a(x,y)=\frac{2a(1-y)+(1-a+ab(x))y}{2-y},
 \qquad a\in\{0,1\},\quad 0\le y\le1.
\end{equation}
For $y_0=1-x$ and $d=y-y_0$ it also satisfies the pointwise
inequality
\begin{equation}\label{eq:F-master}
 F_a(x,y)\le2ax+(1-a)(1-x)+d+3a|d|.
\end{equation}
\end{lemma}

\begin{proof}
Each vertex has candidate inclusion probability $1/Z>0$,
and each round uses an independent sample.  Thus every vertex is
eventually processed with probability $1$.  Hard pairs are separated
by Repulsion and consequently have zero cost in the remaining instance.

For any other unordered pair $e=uv$, set $q=q_{uv}\in[0,1/Z]$.
Until the first round hitting at least one of $u,v$, both vertices
remain unprocessed.  Independence between rounds gives the same
event probabilities in each such round:
\[
 \Prob(\text{exactly one hit})=2/Z-2q,\quad
 \Prob(\text{both hit})=q,\quad
 \Prob(\text{neither hit})=1-2/Z+q.
\]
By conditioning on the first nonempty event for this pair,
its probabilities of being hit separately and simultaneously are
respectively $(2-2Zq)/(2-Zq)$ and $Zq/(2-Zq)$.
In the former case the pair pays boundary cost $a_{uv}$.
In the latter case it is contained in one candidate intersection
$D$. In the upper bound we assign it the charge
$(1-a_{uv})+a_{uv}b(x_{uv})$.
Lemma~\ref{lem:repair} bounds the \emph{sum} of actual internal
costs after Repulsion over all rounds by the sum of these charges.
Taking expectations of that sum and then regrouping by pairs gives
\[
 \E\cost_{I'}(\mathrm{ALG}_\infty)
 \le\sum_{uv\notin H}
 \frac{2a_{uv}(1-Zq_{uv})+
       (1-a_{uv}+a_{uv}b(x_{uv}))Zq_{uv}}{2-Zq_{uv}},
\]
which is \eqref{eq:firsthit-intermediate}.  The bound from Lemma~\ref{lem:repair} applies to the total
internal cost; no separate cut probability is asserted for each pair.

It remains to prove \eqref{eq:F-master}.  Because
$F_a=(1-a)F_0+aF_1$, first consider the two pure signs:
\[
 F_0(x,y)=\frac y{2-y}\le y,
 \qquad
 F_1(x,y)=\frac{2+(b(x)-2)y}{2-y}.
\]
At $y_0=1-x$, the quadratic bound~\eqref{eq:b-quadratic} gives
\[
 F_1(x,1-x)
 =\frac{2x+(1-x)b(x)}{1+x}\le2x.
\]
More precisely, substituting the definition of $b$ gives
\begin{equation}\label{eq:ideal-positive}
 F_1(x,1-x)
 =\min\left\{2x,\frac2{1+x}\right\}\le2x.
\end{equation}
For every fixed $x$, its derivative in $y$ has absolute value
$|2b(x)-2|/(2-y)^2\le2$, since $0\le b\le2$ and $0\le y\le1$.
Therefore
\[
 F_1(x,y)\le2x+2|d|
 \le2x+d+3|d|,
\]
where the last inequality follows separately for $d\ge0$ and
$d<0$.  Similarly, $F_0(x,y)\le y=(1-x)+d$.
Taking their convex combination proves \eqref{eq:F-master}.
\end{proof}

We can now sum the first-hit bound over all pairs. With exact
joint inclusion values, it charges at most twice the positive
part plus once the negative part of the LP objective. The deviations add
the signed error $D_s$ and three times the absolute positive-pair
error $D_+$. The estimates from Section~\ref{sec:lp} control both
terms. The following bound concerns the residual instance $I'$;
the proof of Theorem~\ref{thm:main} restores the fixed costs
removed during preclustering.

\begin{lemma}[Infinite-round objective bound]
\label{lem:rounding-bound}
With $D_s,D_+,Y,P_+$ as defined previously,
\begin{equation}\label{eq:main-objective}
 \E\cost_{I'}(\mathrm{ALG}_\infty)
 \le2L-L_-+D_s+3D_+
 \le2L-L_-+\frac{2Y}{r}+3\sqrt{\frac{2YP_+}{r}}.
\end{equation}
\end{lemma}

\begin{proof}
Sum \eqref{eq:F-master} from Lemma~\ref{lem:firsthit}
over non-hard cross-unit pairs. For a same-unit pair,
$x_{uv}=0$ and $Zq_{uv}=1$, so its contribution to the
first-hit bound is $F_{a_{uv}}(0,1)=1-a_{uv}$, exactly its
LP cost, with no sampling-error term. Adding these contributions
accounts for all non-hard pairs. Since hard pairs in the remaining instance have
$a_{uv}=0$, LP distance $x_{uv}=1$, and hence LP objective zero,
the sum of $2a_{uv}x_{uv}+(1-a_{uv})(1-x_{uv})$
over non-hard pairs is $2L_++L_-=2L-L_-$.
The signed error term is $D_s$,
and the absolute error on positive pairs is $D_+$.
This gives the first inequality in \eqref{eq:main-objective}.
For the second inequality, Lemma~\ref{lem:discrepancy} gives
$D_s\le2Y/r$, and the first inequality of
Lemma~\ref{lem:weighted-error} gives
$D_+\le\sqrt{2YP_+/r}$. Substituting these two bounds yields
\[
 D_s+3D_+\le\frac{2Y}{r}+3\sqrt{\frac{2YP_+}{r}}.\qedhere
\]
\end{proof}

\subsection{Proof of the main theorem}
\label{sec:completion}

We first bound the loss from stopping after finitely many rounds,
then combine this bound with the reductions and the sampling
analysis to prove the main theorem.

Each round selects any fixed vertex with probability $1/Z\ge1/n$.
Thus the chance that it remains unprocessed decreases
exponentially with the number of rounds. The next lemma uses this
fact to stop after $O(n\log(n^2/\tau))$ rounds, paying at most
$\tau$ in additional expected cost. Outputting each remaining
unit separately preserves feasibility.

\begin{lemma}
\label{lem:truncate}
For any rational $\tau\in(0,n^2)$, run
\begin{equation}\label{eq:T-rounds}
 T=n\left\lceil\log_2\frac{n^2}{\tau}\right\rceil
\end{equation}
independent fixed-distribution rounds, then put each unprocessed
unit into its own cluster.  The resulting partition is feasible
and has expected cost in the remaining instance at most the bound in
\eqref{eq:main-objective} plus $\tau$.
\end{lemma}

\begin{proof}
Since each vertex occurs in any independently sampled candidate
with probability $1/Z\ge1/n$ by Lemma~\ref{lem:marginals} \eqref{eq:singleton},
the probability that it remains unprocessed after $T$ rounds is
at most
\[
 (1-1/Z)^T\le e^{-T/n}\le\tau/n^2.
\]
Use the same first $T$ samples to couple this truncated run with
the infinite run.  Pairs with an endpoint processed before time
$T$ have already had their final relation (joined in a repaired
cluster, or separated from the other endpoint) decided, identically
in both runs.  Only pairs with both endpoints still unprocessed
can have different final costs.  The probability that both endpoints of a fixed pair remain is
at most $\tau/n^2$, and any pair contributes
at most one additional unit of cost in the remaining instance.  Summing over
at most $\binom n2<n^2$ pairs proves the $\tau$ additive bound.
The remaining vertices form whole units, which are hard-feasible
when output separately.  Earlier Repulsion steps were hard-feasible,
and splitting further cannot violate any cannot-link constraint.
\end{proof}

We first handle unweighted instances with only cannot-links.
Zero cost can be recognized directly: the positive graph must be
a disjoint union of cliques, and each clique must respect the
cannot-links. After normalization, the latter condition is
automatic. Otherwise the optimum is a positive integer, so a
small additive truncation error is also a small relative error.
Finally, the reduction of
Theorem~\ref{prop:must-only-transfer} handles must-links.

\mainresult*
\begin{proof}
It suffices to consider $0<\varepsilon\le1$: for
$\varepsilon>1$, use the algorithm with $\varepsilon=1$.
We first construct a $(2+\varepsilon)$-approximation for complete
unweighted instances with only cannot-links. We then apply
Theorem~\ref{prop:must-only-transfer} to handle general constrained correlation clustering.

Let $I_0$ be such an unweighted cannot-link-only instance.
Normalize $I_0$ as in Definition~\ref{def:normalization},
obtaining $I$, and let $n=|V(I)|$. Test whether the positive
graph of $I$ is a disjoint union of cliques. If so, its components
give a feasible zero-cost partition of $I$. Restore the removed
vertices as singletons and return. The normalization identity
shows that this partition is optimal for $I_0$, even when its
recorded fixed cost is positive. This also handles the empty
instance.

Otherwise, $\OPT(I)\ge1$. Set $A_0=C_{\mathrm{pre}}+1$ and choose
\begin{equation}\label{eq:parameters}
 r=2\left\lceil\frac{256A_0^2}{\varepsilon^2}\right\rceil,
 \qquad
 \tau=\frac{\varepsilon}{4},\qquad
 T=n\left\lceil\log_2\frac{n^2}{\tau}\right\rceil.
\end{equation}
These choices satisfy $0<\tau<n^2$ and
$\tau\le\varepsilon\OPT(I)/4$. They use only $\varepsilon$, $n$,
and the absolute constant $A_0$, rather than the unknown optimum.

Apply Theorem~\ref{thm:pre}, including its fixed-cost normalization
as in Definition~\ref{def:normalization}. Solve the bounded sub-cluster LP at level $r$ and use the fixed candidate distribution with number of seed steps less than $r/2$.

Lemma~\ref{lem:weighted-error} gives
$Y,P_+\le A_0\OPT(I)$, where $A_0=C_{\mathrm{pre}}+1\ge1$.
Consequently,
\[
 \frac{2Y}{r}+3\sqrt{\frac{2YP_+}{r}}
 \le
 \left(\frac{2A_0}{r}
       +\frac{3\sqrt2 A_0}{\sqrt r}\right)\OPT(I).
\]
The choice in \eqref{eq:parameters} ensures
$r\ge512A_0^2/\varepsilon^2$, so
\[
 \frac{2A_0}{r}\le\frac{\varepsilon^2}{256A_0},
 \qquad
 \frac{3\sqrt2 A_0}{\sqrt r}\le\frac{3\varepsilon}{16}.
\]
Since $0<\varepsilon\le1$ and $A_0\ge1$, their sum is at most
\[
 \frac{\varepsilon}{256}+\frac{3\varepsilon}{16}
 =\frac{49\varepsilon}{256}
 <\frac{\varepsilon}{4}.
\]
Thus the sampling-error contribution is at most
$\varepsilon \OPT(I)/4$.

Lemma~\ref{lem:rounding-bound}, followed by the finite-round
Lemma~\ref{lem:truncate}, bounds the expected cost in the
residual preclustering instance $I'$ by
\[
 2L-L_-+\frac{\varepsilon \OPT(I)}{4}+\tau.
\]
Theorem~\ref{thm:pre} restores the omitted constant cost
$\OPT(I)-\OPT(I')$. Using $L\le\OPT(I')$ (by Lemma~\ref{lem:LP}) and
$\tau\le\varepsilon \OPT(I)/4$ gives expected cost in $I$ at most
\[
 \OPT(I)-\OPT(I')+2L-L_-+\frac{\varepsilon \OPT(I)}{4}+\tau
 \le \OPT(I)+L+\frac{\varepsilon \OPT(I)}{2}
 \le(2+\varepsilon/2)\OPT(I).
\]
Restoring the vertices removed during normalization
as singletons preserves feasibility. The fixed-cost identity for normalization (Definition~\ref{def:normalization}) then gives expected cost at most
$(2+\varepsilon/2)\OPT(I_0)$.

Standard implementations of rational sampling incur an arbitrarily
small additional loss, which we absorb into $\varepsilon$.
All outputs are feasible, and the core algorithm has expected
approximation factor at most $2+\varepsilon$.
Its running time is $|V(I_0)|^{O(\varepsilon^{-2})}$: the LP has
$n^{O(r)}$ variables and constraints, $r=O(\varepsilon^{-2})$,
and the other steps take polynomial time.

Finally, for a general constrained correlation clustering instance, detect infeasibility by
checking whether a cannot-link lies inside a must-link component.
For every feasible input, invoke
Theorem~\ref{prop:must-only-transfer} with the core algorithm
at accuracy $\varepsilon/2$ and transfer slack
$\delta=\varepsilon/2$. The resulting expected approximation
factor is $(2+\varepsilon/2)+\varepsilon/2=2+\varepsilon$.
This completes the algorithm in Figure~\ref{fig:outline}:
Theorem~\ref{prop:must-only-transfer} implements Step~1 and the
projection in the last step; the construction analyzed above
implements preclustering, LP solution, sampling, Repulsion, and
completion. Every output satisfies the original hard constraints.
The expanded vertex count is polynomial in the original count
and $1/\varepsilon$, so the overall running time is
$(n/\varepsilon)^{O(\varepsilon^{-2})}$, now with $n$ denoting
the original input size as in the theorem. The final must-link
projection is deterministic by Lemma~\ref{lem:averaging}.
\end{proof}

\section{Hardness of approximation}\label{sec:hardness}

In this section, we work exclusively on unsigned graphs.
A graph $G$ is \emph{complete} if $V(G)$ is a clique.
A vertex $v$ is \emph{universal} if $N[v] = V(G)$.  
Our reduction is based on adding sufficiently many universal vertices.  This is known as the join of $G$ and a large complete graph in graph-theoretic terminology.

\begin{definition}[Construction]
  Let~$G$ be a graph and let~$t$ be a positive integer.  The graph~$\Gt$ is obtained from~$G$ by adding a clique~$U$ of~$t$ new vertices and making every vertex of~$U$ adjacent to every vertex of~$V(G)$.
\end{definition}

Let~$\OPT(G)$ denote the size of a minimum cluster deletion set of~$G$, and let $\omega(G)$ denote the size of a maximum clique of~$G$.

\begin{proposition}\label{lem:vc-ce}
  Let~$n = |V(G)|$ and~$m = |E(G)|$.  If $m > 0$, then
  \[
    \OPT(\Gt) \le 
      t(n-\omega(G)) + m - \binom{\omega(G)}{2}
      \le
      t(n-\omega(G)) + m.
    \]
    Moreover, if $m>0$, then the second inequality is strict.
\end{proposition}

\begin{proof}
  Let $H = \Gt$.
  Let~$S$ be a maximum clique of~$G$, and let $V_{-} = V(G)\setminus S$.
  Delete all edges of $H$ incident to vertices of $V_-$.  The resulting graph is a cluster graph: one cluster is $U\cup S$, and every vertex of $V_-$ is an isolated singleton cluster.
  The cost of this solution is
  \[
    |U| \cdot |V_{-}| + |E({G})\setminus E({G}[S])| =
    t (n - \omega(G)) + m - \binom{|S|}{2}
    = t(n-\omega(G)) + m - \binom{\omega(G)}{2}.
\]
  It is strictly smaller than $t (n - \omega(G)) + m$ because~$\omega(G) > 1$ when $m > 0$.
\end{proof}

By construction, the vertices of $U$ are universal true twins in
$\Gt$. The following is a direct consequence of
Theorem~\ref{thm:weighted-twins}.

\begin{corollary}
\label{lem:true-twins}
Every minimum cluster deletion set preserves every pair of true
twins. Moreover, any cluster deletion set can be transformed in
polynomial time, without increasing its size, into one preserving
every pair of true twins.
\end{corollary}
\begin{proof}
Model cluster deletion as constrained correlation clustering: graph edges have separation cost
one and co-clustering cost zero, while graph nonedges are
cannot-links. True twins have identical external costs and
cannot-link relations. Theorem~\ref{thm:weighted-twins}
therefore implies that every optimal partition keeps them together.
For an arbitrary solution, whenever true twins are separated,
apply its improving move. Each move decreases the integer cost
by at least one, so after at most $|E(G)|$ moves no true-twin
pair is separated. Each move can be found and performed in
polynomial time.
\end{proof}

\begin{lemma}\label{lem:ce-vc}
  Let~$n=|V(G)|$.  Given any cluster deletion set~$E_-$ of~$\Gt$, one can produce, in time $O((t+n)^2)$, a clique of~$G$ of size at least
  \[
    n-\frac{|E_-|}{t}.
  \]
\end{lemma}
\begin{proof}
  Let~$H=\Gt$, and let~$U$ be the set of the $t$ added universal vertices.
  Among the clusters of $H-E_-$ meeting $U$, choose $C$ maximizing
  $|C\cap V(G)|$ and gather all of $U$ into it. This preserves
  feasibility, does not increase the separation cost between $U$
  and $V(G)$, and removes all separation cost within $U$.
  Let $E'_-$ be the resulting deletion set; then $|E'_-|\le|E_-|$.
  The cluster $C$ can be chosen and the gathering performed in
  $O((t+n)^2)$ time, as in the projection of Lemma~\ref{lem:averaging}.

  We return the clique
  \[
    K = C\setminus U.
  \]
  Because $C$ is a clique in $H-E'_-$ and no edge inside $V(G)$ is introduced, $K$ is a clique of $G$.
  Every vertex in~$V(G)\setminus K$ lies outside the cluster containing~$U$.
  Therefore, all edges between~$U$ and~$V(G)\setminus K$ must be deleted by
  $E'_-$.  Hence
  \[
    |E'_-|
    \ge
    |U|\cdot |V(G)\setminus K|
    =
    t(n-|K|).
  \]
  Since~$|E'_-|\le |E_-|$, we obtain
  \[
    |K|
    \ge
    n-\frac{|E'_-|}{t}
    \ge
    n-\frac{|E_-|}{t}.
  \]

  Given $E_-$ explicitly, the clustering of $H-E_-$ and the set $K$ can be computed in $O(|V(H)|^2)$ time.
\end{proof}

Proposition~\ref{lem:vc-ce} and Lemma~\ref{lem:ce-vc} already imply the NP-hardness of cluster deletion: it suffices to set~$t= |E(G)|$.
For the main theorem, we use the following standard relation between cliques and vertex covers.  The \emph{complement graph}~$\overline{G}$ of a graph~$G$ is defined on the same vertex set~$V(G)$, where two distinct vertices~$u$ and~$v$ are adjacent in~$\overline{G}$ if and only if~$uv\notin E(G)$.
Note that the complement of~$\overline{G}$ is~$G$.
Let~$\tau(G)$ denote the size of a minimum vertex cover of~$G$.  Since a clique in a graph is an independent set in its complement, we have
\[
  \tau(G) = |V(G)| - \omega \left(\overline{G} \right).
\]

\vctransfer*
\begin{proof}
  Let~$G$ be the input graph for vertex cover, and let~$n=|V(G)|$.  If~$G$ is edgeless, then the empty set is an optimal vertex cover.  If~$G$ is complete, then any set of~$n-1$ vertices is an optimal vertex cover.  Thus we may assume that~$G$ has at least one edge and is not complete.

  Choose an arbitrary edge~$v_1v_2\in E(G)$.  For~$i=1,2$, define
  \[
    G_{i} = {G - v_{i}} \quad \text{ and }\quad
    H_{i} = \overline{G_{i}} \vee K_{t},
  \]
  where 
  \[
t = \left\lceil \frac{c n^{2}}{2(c-1)} \right\rceil.
  \]
  Since every vertex cover of~$G$ contains at least one endpoint of the edge~$v_1v_2$, there is an index~$i\in\{1,2\}$ such that deleting~$v_i$ from a minimum vertex cover of~$G$ leaves a vertex cover of~$G_i$.  Hence
  \[
    \min_{i \in \{1, 2\}}  \tau(G_{i}) \le \tau(G) - 1.
  \]

  For each~$i\in\{1,2\}$, run the assumed $c$-approximation algorithm for cluster deletion on~$H_i$, and let~$E_i$ be the returned deletion set.  By the approximation guarantee and Proposition~\ref{lem:vc-ce},
  \[
    |E_{i}| \le c\cdot \OPT(H_{i}) \le c \left( t \left( |V(G_i)| -\omega \left(\overline{G_{i}} \right) \right) + \left| E(\overline{G_{i}})\right| \right) < c t \tau(G_{i}) + c n^{2} / 2.
  \]
  We may assume without loss of generality that $|E_{1}| \le |E_{2}|$.  Then
  \[
|E_{1}|
    = \min_{i \in \{1, 2\}} |E_{i}|
    < c t \cdot \min_{i \in \{1, 2\}} \tau(G_{i}) + \frac{c n^2}{2} \le c t \left( \tau(G) - 1 \right) + \frac{c n^2}{2}.
  \]
  Since~$c n^{2} / 2 \le t(c-1)$ by the choice of~$t$, we have
  \[
    |E_{1}| < c t (\tau(G) - 1) + t(c-1) = c t\, \tau(G) - t.
  \]

  Now apply Lemma~\ref{lem:ce-vc} to~$E_1$ in~$H_1$.  This yields a clique $K$ of~$\overline{G_1}$ such that
  \[
    |K| \ge \left| V \left( \overline{G_{1}} \right) \right| - \frac{|E_{1}|}{t} \ge n - 1 - \frac{c t\, \tau(G) - t}{t} = n - c\tau(G).
  \]
  Since~$K$ is a clique in~$\overline{G_1}$, it is an independent set in~$G_1$ and~$G$.  Hence, $V(G)\setminus K$ is a vertex cover of~$G$.
  The size of this vertex cover is
  $
    |V(G)\setminus K|
    =
    n-|K|
    \le c\tau(G).
  $
  Thus, we obtain a $c$-approximation for vertex cover.

  It remains to analyze the running time.  Let~$c' = \left\lceil \frac{c}{2(c-1)} \right\rceil + 1.$  Then
  each graph~$H_i$ has
  \[
    |V(H_i)| = n-1+ t \le c' n^2
  \]
  vertices.  Constructing~$\overline{G_i}$ takes $O(n^2)$ time, and explicitly constructing the join with~$K_t$ takes
  $
    O(|V(H_i)|^2)=O(n^4)
  $
  time.
  The assumed approximation algorithm is invoked twice, each time on a graph with at most~$c'n^2$ vertices.  Thus, the total time spent in the approximation algorithm is $2 f(c'n^2)$.  Finally, applying Lemma~\ref{lem:ce-vc} takes
  $ O((t+n)^2)=O(n^4) $ time.  Therefore, the total running time is
  \[
    O\left( f(c' n^{2}) + n^4 \right),
  \]
  as claimed.
\end{proof}

\hardnessresult*
\begin{proof}
The vertex cover hardness following from Khot et al.~\cite{khot-23-ugc} rules out a $(\sqrt{2}-\varepsilon)$-approximation
unless $\mathrm P=\mathrm{NP}$. Khot and Regev~\cite{khot-08-vc-ugc}
established $(2-\varepsilon)$-inapproximability assuming the Unique
Games Conjecture. The approximation-preserving reduction of
Theorem~\ref{thm:reduction} transfers both bounds to cluster
deletion, and hence to constrained correlation clustering with
only cannot-links.
\end{proof}

The reduction also explains the algorithmic correspondences in
Table~\ref{tab:vc-correspondences}. Write the constructed graph as
$G=\overline Q\vee U$, where $Q$ is the vertex cover instance and
$U$ is the added universal clique. After gathering $U$ into one
cluster, the vertices kept with it form an independent set of
$Q$; the vertices separated from it form a vertex cover.
For the metric-LP connection, fix $u\in U$ and put $z_v=x_{uv}$.
Every edge $vw\in E(Q)$ is a non-edge in $G$, so the triangle
inequality gives $z_v+z_w\ge1$: these are the vertex-cover LP
constraints. The STC constraints on the edges incident to this
fixed $u$ give the same inequalities, because $u,v,w$ induce a
three-vertex path in $G$ exactly when $vw\in E(Q)$. The table
follows the choices involving $U$; the clustering algorithms also
process the vertices separated from it.

\begin{table}[htp]
  \centering
  \renewcommand{\arraystretch}{1.18}
  \caption{Cluster-deletion methods on $G=\overline Q\vee U$
    and their vertex-cover interpretations on $Q$.}
  \label{tab:vc-correspondences}
  \begin{tabularx}{\textwidth}{@{}
    >{\raggedright\arraybackslash}p{.34\textwidth}
    >{\raggedright\arraybackslash}X@{}}
    \toprule
    Cluster-deletion method & Vertex-cover interpretation \\
    \midrule
    Maximum-clique removal \cite{dessmark-07-edge-clique-partition}
                            & Complement of a maximum independent set (NP-hard). \\
    \addlinespace[7pt]
    Metric-ball rounding    \cite{charikar-05-approximate-clustering,puleo-15-bounded-cluster-sizes}
                            & Threshold distances $z_v=x_{uv}$ from $u\in U$; vertices outside the resulting feasible cluster form the cover. \\
    \addlinespace[7pt]
    CKR partitions
    \cite{adriaens-26-ckr-partitions}
                            & Use randomized metric partitioning; vertices outside the part containing $U$ form the cover. \\
    \addlinespace[7pt]
    STC-LP and Pivot \cite{veldt-22-clustering-via-strong-triadic-closure,balmaseda-24-cluster-deletion} & Round the half-integral vertex-cover LP. \\
    \addlinespace[7pt]
    MatchFlipPivot \cite{veldt-22-clustering-via-strong-triadic-closure,balmaseda-24-cluster-deletion}
                            & Find a maximal matching and take all its endpoints.\\
    \midrule
    \addlinespace[2pt]
    Repulsion
                            & Complement of a maximal independent set. \\
    \bottomrule
  \end{tabularx}
\end{table}

\section{Concluding remarks}\label{sec:dis}

We discuss the remaining obstacles to derandomization and an exact
factor-two guarantee, and the scope of possible extensions to
weighted instances.

Consider the algorithm in Figure~\ref{fig:outline}.
The metric-LP initialization used in Step~2 and Repulsion in
Step~5.2 admit deterministic center choices. The must-link
projection in Step~7 is also deterministic by
Lemma~\ref{lem:averaging}. Randomness remains in the
fractional-weight reduction in Step~1, including its choice of
copy representatives when projecting back in Step~7, and in
candidate sampling in Step~5.1. The main obstacles to
derandomization are the simultaneous concentration bound over
all partitions in the expansion analysis and the averaging over
seed histories and stopping times in the sampling analysis.

The approximation slack comes from the reduction in Step~1,
sampling error in Step~5.1, and stopping the repeated rounds in
Step~5. The sampling bound contains an $O(r^{-1/2})$ term, so we
choose the LP level in Step~3 to be
$r=\Theta(\varepsilon^{-2})$. Setting $\varepsilon=0$ therefore
does not yield an exact factor-two algorithm from these estimates.
Such a guarantee would require eliminating the losses or charging
them against additional savings.

\paragraph{Two-weight correlation clustering.}

After setting aside fixed costs, hard constraints can be encoded
by assigning a sufficiently large disagreement penalty to the
corresponding preferences. This places constrained correlation
clustering within weighted correlation clustering with two
weights, $1$ and $W$. On feasible constrained instances, this
encoding can preserve approximation guarantees when $W$ is
chosen sufficiently large. It does not, however, allow us to
apply our algorithm to arbitrary two-weight instances.

To see the distinction, suppose the high-weight preferences are
jointly satisfiable. If $W>\binom n2$, a partition satisfying all
of them costs less than $W$, whereas violating even one costs at
least $W$. Every optimum therefore respects these preferences,
and the instance can be treated as constrained correlation
clustering. More generally, choosing $W>\alpha\binom n2 > \alpha \OPT$ forces
every $\alpha$-approximate partition to satisfy all high-weight
preferences. If these preferences are inconsistent, treating them
as hard constraints instead produces an infeasible instance,
although the finite-weight problem still has a finite optimum.

In the inconsistent case, some high-weight disagreement is
unavoidable. After dividing costs by $W$, the high-weight pairs
have weight one and the remaining pairs have weight $1/W$.
Omitting the latter changes every partition cost by at most
$\binom n2/W$. Thus, for sufficiently large $W$, two-weight
instances can approximate correlation clustering on incomplete
signed graphs with arbitrarily small relative loss. The following
proposition formalizes this observation and extends the transfer
to general weighted instances.

\begin{proposition}\label{prop:two-weight-transfer}
Fix constants $\alpha\ge1$ and $\delta>0$.
A polynomial-time $\alpha$-approximation for complete signed
correlation clustering with disagreement weights in $\{1,W\}$,
where $W$ is part of the input, gives a polynomial-time
$(\alpha+\delta)$-approximation for general weighted
correlation clustering.
\end{proposition}

\begin{proof}
First consider unweighted correlation clustering on an incomplete
signed graph $G$ with $n$ vertices. Give every specified pair its
original sign and weight $W$, and every unspecified pair an
arbitrary sign and weight one. For every partition $\calP$,
\[
 W\,\cost_G(\calP)
 \le\cost_W(\calP)
 \le W\,\cost_G(\calP)+\binom n2.
\]
An $\alpha$-approximate partition of the completed instance
therefore satisfies
\[
 \cost_G(\calP)
 \le\alpha\OPT(G)+\frac{\alpha\binom n2}{W}.
\]
Zero cost is possible exactly when no negative edge has both
endpoints in one connected component of the positive-edge graph;
this can be checked directly. Otherwise $\OPT(G)\ge1$, and choosing
\[
 W=\max\left\{2,\left\lceil
       \frac{2\alpha\binom n2}{\delta}\right\rceil\right\}
\]
gives a $\beta$-approximation for the incomplete problem, where
$\beta=\alpha+\delta/2$.

We next reduce general nonnegative rational weights to polynomially
bounded integer weights, allowing additional slack $\eta=\delta/2$.
Discard zero-weight pairs and handle zero-cost instances directly.
Let $m$ be the number of remaining pairs. Using a polynomial-time
$\rho$-approximation, with polynomially bounded $\rho$, obtain
a partition of cost $B>0$; such an algorithm is available
by~\cite{demaine-06-correlation-clustering}. Then $B/\rho\le\OPT\le B$. Put
\[
 h=\frac{\eta B}{\beta\rho m},\qquad
 K=\left\lceil\beta\left(\frac Bh+m\right)\right\rceil+1,
 \qquad
 w'_e=\min\{\lceil w_e/h\rceil,K\}.
\]
The integer weights are polynomially bounded. The partition of
cost $B$ has rounded cost at most $B/h+m$, so a
$\beta$-approximate rounded solution has cost strictly less than
$K$ and cannot disagree with any pair whose weight was capped.
Its original cost is consequently at most
\[
 h\beta\OPT_{w'}
 \le\beta\OPT+\beta mh
 \le(\beta+\eta)\OPT.
\]

To obtain the required approximation for integer weights,
simulate each pair $uv$ by $w'_e$ internally disjoint length-two
paths with fresh internal vertices. Use two positive edges for a
positive pair, and one positive and one negative edge for a
negative pair; all other pairs remain unspecified. For fixed
endpoint clusters, each path has minimum cost zero when the
original preference is satisfied and one otherwise. Thus every
original partition extends at equal cost, and restricting any
expanded partition to the original vertices cannot increase cost.
Applying the incomplete-graph algorithm to this polynomial-size
expansion gives the required
$(\beta+\eta)=(\alpha+\delta)$ guarantee.
\end{proof}

Correlation clustering on incomplete graphs and multicut are
approximation-equivalent up to constant factors~\cite{demaine-06-correlation-clustering}.
Together with the UGC-hardness of approximating multicut within
any constant~\cite{chawla-06-hard-multicut},
Proposition~\ref{prop:two-weight-transfer} rules out a
constant-factor approximation for unrestricted two-weight
instances under the same assumption.

Our result leaves open whether the $(2+\varepsilon)$ guarantee extends to instances with hard constraints and disagreement weights in $\{1,c\}$ for fixed $c>1$. Better bounds are known without hard constraints. For $1\le c\le2$, the disagreement weights satisfy the triangle inequality, so the $(2+\varepsilon)$-approximation of Fan et al.~\cite{fan-26-pseudometric-weighted-correlation-clustering} applies. For $c>2$, replace every weight $c$ by $2$ and apply their algorithm. This decreases every partition cost by a factor of at most $c/2$, giving a $(c+\varepsilon)$-approximation after rescaling the accuracy parameter.

\paragraph{Finding the intersection of all optimal partitions.}
Define two vertices to be equivalent when every optimal partition
keeps them together. The equivalence classes form the common
refinement of all optimal partitions. Our atoms are certified
subsets of these classes, rather than the entire classes.
The following lemma explains why we do not try to compute all
of them.

\begin{lemma}\label{lem:all-preserved-sets}
For unweighted cluster deletion, deciding whether every optimal
partition keeps two specified vertices together is coNP-hard.
\end{lemma}
\begin{proof}
  Start from a clique instance $(Q,k)$, with $1\le k\le |V(Q)|$.  Obtain $Q'$ by replacing each vertex of $Q$ with two adjacent true twins and add a disjoint clique of $2k-1$ vertices.  Then
  \[
    \omega(Q') = \max\{ 2\omega(Q), 2k - 1\}.
  \]
  Let
  \[
    G = Q'\vee K_t,\qquad t=|E(Q')|+1,
  \]
and let $U$ be the added universal clique.  True-twin preservation (Corollary~\ref{lem:true-twins}) keeps all of $U$ in one cluster in every optimum of $G$. The vertices of that cluster outside $U$ form a maximum clique of $Q'$: otherwise the separation cost from $U$ alone exceeds $t(|V(Q')|-\omega(Q'))+|E(Q')|$, the cost upper bound obtained by putting $U$ with a maximum clique of $Q'$ and leaving all other vertices as singletons.

Choose $u\in U$ and $v$ in the added clique of $Q'$. Then every optimal partition keeps $u$ and $v$ together if and only if
\[
 2\omega(Q)<2k-1
 \quad\Longleftrightarrow\quad
 \omega(Q)<k.
\]
Thus the construction reduces the complement of Clique to the stated problem.
\end{proof}

By Lemma~\ref{lem:all-preserved-sets}, membership of this decision problem in NP would imply $\mathrm{NP}=\mathrm{coNP}$.

For unconstrained correlation clustering, the analogous decision problem is NP-hard under polynomial-time Turing reductions. Indeed, an oracle for it would let us reduce the optimum by one at each step: on an induced path $u-v-w$, delete $uv$ or $vw$ if some optimum separates its endpoints; otherwise every optimum joins all three vertices, and we add $uw$. In either case the optimum decreases by exactly one. Repeating until the graph is a disjoint union of cliques computes the original optimum in polynomially many oracle calls. 

\appendix

\section[Radius-2/3 Repulsion]{Radius-$2/3$ Repulsion: Theorem~\ref{thm:metric-repulsion}}
\label{sec:metric-repulsion}

We prove Theorem~\ref{thm:metric-repulsion}, illustrated in Figure~\ref{fig:metric-repulsion-example}.
Call a pair \emph{short} if its distance is less than $2/3$.
For a current set $S\subseteq V$ and a center $c\in S$, define
the strict ball
\[
 A_c=\{u\in S:x_{cu}<2/3\}.
\]
Since $x_{cc}=0$, this ball is nonempty.
Separating $A_c$ from $S\setminus A_c$ incurs boundary cost
$e^+(A_c,S\setminus A_c)$.

To analyze the algorithm, assign each unordered pair the
nonnegative potential
\[
 \phi_{uv}=
 \begin{cases}
  3x_{uv},&uv\in E^+,\ x_{uv}<2/3,\\
  1,&uv\in E^+,\ x_{uv}\ge2/3,\\
  1,&uv\in E^-,\ x_{uv}<2/3,\\
  0,&uv\in E^-,\ x_{uv}\ge2/3,
 \end{cases}
 \qquad
 \Phi(S)=\sum_{uv\in\binom S2}\phi_{uv}.
\]
For a short pair, put
\[
 h_{uv}=
 \begin{cases}
  1-3x_{uv},&uv\in E^+,\\
  -1,&uv\in E^-.
 \end{cases}
\]
The boundary cost plus the change in potential is
\begin{align}
 \Delta_c
 &=e^+(A_c,S\setminus A_c)
     +\Phi(A_c)+\Phi(S\setminus A_c)-\Phi(S)\notag\\
 &=\sum_{\substack{u\in A_c,\ v\in S\setminus A_c\\
                   x_{uv}<2/3}}h_{uv}.
 \label{eq:pivot-excess}
\end{align}
Indeed, a short positive boundary pair contributes $1-3x_{uv}$,
and a short negative boundary pair contributes $-1$.
A non-short boundary pair contributes zero, since its potential
equals its separation cost.
Choosing centers uniformly at random gives the expected bound; choosing a center with nonpositive potential excess gives its deterministic version.
\metricrepulsion*
\begin{proof}
We first bound the initial potential by the LP objective.
For a positive pair, $\phi_{uv}\le3x_{uv}$:
there is equality for a short pair, and otherwise
$\phi_{uv}=1\le3x_{uv}$.
For a negative pair, $\phi_{uv}\le3(1-x_{uv})$:
a short pair has potential one and $x_{uv}<2/3$,
while a non-short pair has potential zero.
Therefore
\begin{equation}\label{eq:repulsion-initial-potential}
 \Phi(V)\le3L(x).
\end{equation}

We next show that a uniformly chosen proper ball has
nonpositive expected $\Delta_c$.
Form the auxiliary graph on $S$ whose edges are the short pairs.
Its closed neighborhoods are exactly the balls $A_c$.
Sum \eqref{eq:pivot-excess} over all centers $c\in S$.

Every term in this sum corresponds to an induced three-vertex
path in the auxiliary graph.
For an induced path $u-v-w$, the center $u$ contributes
$h_{vw}$ and the center $w$ contributes $h_{uv}$.
For every short pair,
\[
 h_{uv}\le1-3x_{uv};
\]
for a negative pair, this follows from $x_{uv}<2/3$.
The combined contribution of the path is therefore at most
\[
 h_{uv}+h_{vw}
 \le2-3(x_{uv}+x_{vw})
 \le2-3 x_{uw}
 \le0,
\]
where the second inequality uses the triangle inequality and the last holds because the missing pair $uw$ is not short.
Grouping all contributions by these induced paths gives
\[
 \sum_{c\in S}\Delta_c\le0.
\]
Centers with $A_c=S$ contribute zero. Hence, whenever a split
is needed, the uniform choice among proper centers satisfies
\[
 \E_c\!\left[
 e^+(A_c,S\setminus A_c)
 +\Phi(A_c)+\Phi(S\setminus A_c)
 \right]\le\Phi(S).
\]

We prove by induction on $|S|$ that the expected cost of the
recursive partition of $S$ is at most $\Phi(S)$.
For a terminal block, all pairs are short.
Its internal disagreement cost is the number of negative pairs,
each of which has potential one, so its cost is at most $\Phi(S)$.
For a nonterminal block, condition on the chosen center.
By induction, the expected costs of the two recursive calls
are at most their respective potentials.
Adding the boundary cost and averaging over the center gives
the required bound for $S$.
At the root, \eqref{eq:repulsion-initial-potential} yields
\[
 \E[\cost_I(\calP_0)]\le\Phi(V)\le3L(x).
\]

For feasibility, vertices at distance zero have identical
distances to every center and are never separated by a split.
In particular, all must-links are preserved.
Every terminal block contains only pairs at distance less than
$2/3$, so it contains no cannot-link, whose distance is one.

Replace the uniform random choice of a center defining a proper
ball by the following rule:
\[
 \text{choose }c\in\arg\min_{\substack{c\in S\\A_c\ne S}}\Delta_c,
 \quad\text{breaking ties deterministically}.
\]
The chosen split then satisfies
\[
 e^+(A_c,S\setminus A_c)
 +\Phi(A_c)+\Phi(S\setminus A_c)
 \le\Phi(S).
\]
This inequality holds at every split, rather than only in
expectation. Induction over the recursion therefore gives
\[
 \cost_I(\calP_0)\le\Phi(V)\le3L(x).
\]
Computing all candidate values $\Delta_c$ takes polynomial time,
and the feasibility argument is unchanged.

Finally, every proper split produces two nonempty parts.
There are at most $|V|-1$ splits, and each center and ball can
be evaluated in polynomial time.
\end{proof}

\section[Probability-weighted expansion]{Probability weights and constraint-preserving expansion}
\label{app:probability-transfer}

We prove Theorem~\ref{prop:constraint-transfer}, extending the random expansion of Chawla et al.~\cite[Section~6, Theorem~20]{chawla-15-correlation-clustering} to cannot-link constraints.\footnote{Their theorem is stated with
factor $\alpha$, while the sketch allows an additive error.
We make the arbitrarily small slack in the approximation ratio explicit.} With no cannot-links, this is the usual transfer from probability-weighted to unweighted correlation clustering.

\probabilitytransfer*

\begin{proof}
  Let $I$ be a probability-weighted instance on $n$ vertices
  with cannot-link set $H$, separation costs $p_{uv}$, and
  co-clustering costs $1-p_{uv}$. Instances with fewer than two
  vertices are immediate. Write $\varepsilon=\min\{\delta,1\}$,
  where $\delta$ is the slack in the proposition, and put
  $m=\binom n2$.

  First handle instances whose optimal value may be arbitrarily small.
  Give each pair its cheaper relation: label it positive if
  $p_{uv}\ge1/2$ and negative otherwise.
  If the resulting positive graph is a disjoint union of cliques
  and its component partition respects $H$, return that partition.
  It minimizes every pair contribution simultaneously and has cost
  \[
    \sum_{u<v}\min\{p_{uv},1-p_{uv}\}.
  \]
  Otherwise every feasible partition disagrees with at least one
  majority label: either no partition satisfies all these labels,
  or their unique component partition violates a cannot-link.
  Such a disagreement costs at least $1/2$, so
  \begin{equation}\label{eq:probability-opt-lower}
    \OPT(I)\ge\frac12.
  \end{equation}
  Henceforth assume this second case.

  Set
  \[
    \gamma=\frac{\varepsilon}{8(\alpha+1)},
    \qquad
    \eta=\frac{\varepsilon}{8m}.
  \]
  Replace each vertex $v$ by a set $V_v$ of $T$ copies.
  Make every pair within $V_v$ positive. For copies of distinct
  vertices $u,v$, independently label their pair positive with
  probability $p_{uv}$ and negative with probability $1-p_{uv}$.
  For every $uv\in H$, impose a cannot-link on every pair in
  $V_u\times V_v$. Denote this set by $H_T$. This gives a
  complete unweighted random instance $I_T$ on $N=nT$ vertices,
  with only cannot-links. No must-links are imposed within groups
  of copies. In the application after averaging, $p_{uv}=0$
  for $uv\in H$, so these pairs also have deterministic negative
  labels; the proof permits arbitrary probability costs on $H$.

  For every fixed partition $\calQ$ of the copies, define
  \[
    \overline c(\calQ)=\E[\cost_{I_T}(\calQ)].
  \]
  This is a deterministic function of $\calQ$: it is the cost
  obtained by using the original probability weights between
  different groups of copies and unit positive weights within
  each group.

  \emph{Uniform concentration.}
  For a fixed $\calQ$, its random cost is the sum of at most
  $N^2/2$ independent zero--one variables and a deterministic
  term. Hoeffding's inequality gives
  \[
    \Prob\!\left[
      |\cost_{I_T}(\calQ)-\overline c(\calQ)|>\gamma T^2
    \right]
    \le
    2\exp\!\left(-\frac{4\gamma^2T^2}{n^2}\right).
  \]
  There are at most $N^N$ partitions. Consequently,
  \[
    \Prob\!\left[
      \exists\calQ:
      |\cost_{I_T}(\calQ)-\overline c(\calQ)|>\gamma T^2
    \right]
    \le
    2\exp\!\left(
      nT\log(nT)-\frac{4\gamma^2T^2}{n^2}
    \right).
  \]
  Choose an integer $T$ large enough that this bound is at most
  $\eta$. For example, the polynomial-size choice
  \[
    L=\left\lceil\log\frac{2n}{\gamma\eta}\right\rceil,\qquad
    T=\left\lceil256n^4\gamma^{-4}L^2\right\rceil
  \]
  suffices: using $T\le512n^4\gamma^{-4}L^2$ gives
  $\log(nT)\le18L$, while the magnitude of the negative term
  is at least $1024Tn^2\gamma^{-2}L^2$.
  Call the complementary event the good event.

  The estimate holds simultaneously for all partitions.
  It therefore applies to the partition returned by an algorithm
  after observing the random labels.

  \emph{Bounding the expanded optimum.}
  Lift an optimal feasible partition of $I$ by placing all copies
  of each original vertex in its cluster. This respects $H_T$,
  because every pair in $H$ is separated in the original partition.
  The expected cost of this lifted partition is
  $T^2\OPT(I)$. On the good event, its actual cost is at most
  $T^2\OPT(I)+\gamma T^2$. Thus
  \[
    \OPT(I_T)\le T^2\OPT(I)+\gamma T^2.
  \]

  \emph{Projection to the original vertices.}
  Run the assumed algorithm for unweighted instances with only
  cannot-links on $I_T$, obtaining a feasible partition $\calQ$.
  Independently for each original vertex $v$, choose a uniformly
  random representative from $V_v$ and assign $v$ its
  representative's cluster label. Denote the resulting partition
  by $\calP$. This partition is feasible: if $uv\in H$, every
  possible choice of their two representatives is a pair in $H_T$,
  and hence the representatives belong to different clusters.
  The requirement to forbid all $T^2$ cross-group pairs is what
  makes this implication hold.

  For every fixed $\calQ$,
  \begin{equation}\label{eq:copy-projection}
    \E[\cost_I(\calP)\mid\calQ]
    \le \frac{\overline c(\calQ)}{T^2}.
  \end{equation}
  Indeed, for each distinct original pair $u,v$, its expected
  contribution is the average probability-weighted contribution
  over the $T^2$ pairs in $V_u\times V_v$.
  Summing gives the cross-group part of
  $\overline c(\calQ)/T^2$.
  The remaining contributions, from separating copies within
  one group, are nonnegative and disappear upon projection.

  Fix a realization of $I_T$ satisfying the good event.
  Uniform concentration, the approximation guarantee, and the
  expanded-optimum bound imply
  \begin{align*}
    \E[\cost_I(\calP)\mid I_T]
    &\le \frac{\E[\overline c(\calQ)\mid I_T]}{T^2}\\
    &\le \frac{\E[\cost_{I_T}(\calQ)\mid I_T]}{T^2}
      +\gamma\\
    &\le \frac{\alpha\OPT(I_T)}{T^2}+\gamma\\
    &\le \alpha\OPT(I)+(\alpha+1)\gamma.
  \end{align*}
  Here the expectations include the unweighted algorithm's
  randomness, if any, and the representative choices.

  On the bad event, the projected partition still costs at
  most $m$, because each original pair contributes at most one.
  Averaging over the random expansion therefore gives
  \[
    \E[\cost_I(\calP)]
    \le \alpha\OPT(I)+(\alpha+1)\gamma+\eta m
    =\alpha\OPT(I)+\frac{\varepsilon}{4}.
  \]

  Using \eqref{eq:probability-opt-lower}, the algorithm satisfies
  \[
    \E[\cost_I(\calP)]
    \le \alpha\OPT(I)+\frac{\varepsilon}{4}
    \le(\alpha+\varepsilon)\OPT(I)
    \le(\alpha+\delta)\OPT(I).
  \]
  The expansion has polynomial size, and the construction and
  projection always preserve feasibility. 
\end{proof}


\begin{thebibliography}{10}

\bibitem{adriaens-26-ckr-partitions}
Florian Adriaens.
\newblock {CKR} partitions and lower bounds for constrained correlation
  clustering and variants, 2026.
\newblock \href {https://doi.org/10.48550/ARXIV.2608.08291}
  {\path{doi:10.48550/ARXIV.2608.08291}}.

\bibitem{adriaens-26-bad-triangle-transversals}
Florian Adriaens and Nikolaj Tatti.
\newblock Simple algorithms for bad triangle transversals with applications to
  correlation clustering.
\newblock In \emph{ICML}, PMLR 306:675--689. PMLR, 2026.
\newblock URL: \url{https://proceedings.mlr.press/v306/adriaens26a.html}.

\bibitem{ailon-08-aggregating-inconsistent-information}
Nir Ailon, Moses Charikar, and Alantha Newman.
\newblock Aggregating inconsistent information: Ranking and clustering.
\newblock {\em Journal of the ACM}, 55(5):(Article 23) 1--27, 2008.
\newblock \href {https://doi.org/10.1145/1411509.1411513}
  {\path{doi:10.1145/1411509.1411513}}.

\bibitem{azizeddin-26-constrained-correlation-clustering}
Sina Azizeddin, Evangelos Kipouridis, and Nithin Varma.
\newblock Constrained correlation clustering: Towards optimality, 2026.
\newblock \href {https://doi.org/10.48550/ARXIV.2608.05700}
  {\path{doi:10.48550/ARXIV.2608.05700}}.

\bibitem{balmaseda-24-cluster-deletion}
Vicente Balmaseda, Ying Xu, Yixin Cao, and Nate Veldt.
\newblock Combinatorial approximations for cluster deletion: Simpler, faster,
  and better.
\newblock In \emph{ICML}, PMLR 235:2586--2606, 2024.
\newblock URL: \url{https://proceedings.mlr.press/v235/balmaseda24a.html}.

\bibitem{bansal-04-correlation-clustering}
Nikhil Bansal, Avrim Blum, and Shuchi Chawla.
\newblock Correlation clustering.
\newblock {\em Machine Learning}, 56(1):89--113, 2004.
\newblock \href {https://doi.org/10.1023/B:MACH.0000033116.57574.95}
  {\path{doi:10.1023/B:MACH.0000033116.57574.95}}.

\bibitem{cao-22-edge-modification}
Gabriel Bathie, Nicolas Bousquet, Yixin Cao, Yuping Ke, and Th\'{e}o Pierron.
\newblock ({S}ub)linear kernels for edge modification problems toward
  structured graph classes.
\newblock {\em Algorithmica}, 84:3338--3364, 2022.
\newblock \href {https://doi.org/10.1007/s00453-022-00969-1}
  {\path{doi:10.1007/s00453-022-00969-1}}.

\bibitem{cao-25-sublinear}
Nairen Cao, Vincent Cohen{-}Addad, Euiwoong Lee, Shi Li, David~Rasmussen Lolck,
  Alantha Newman, Mikkel Thorup, Lukas Vogl, Shuyi Yan, and Hanwen Zhang.
\newblock Solving the correlation cluster {LP} in sublinear time.
\newblock In \emph{STOC}, pages 1154--1165, 2025.
\newblock \href {https://doi.org/10.1145/3717823.3718181}
  {\path{doi:10.1145/3717823.3718181}}.

\bibitem{cao-24-lp}
Nairen Cao, Vincent Cohen{-}Addad, Euiwoong Lee, Shi Li, Alantha Newman, and Lukas Vogl.
\newblock Understanding the cluster linear program for correlation clustering.
\newblock In  \emph{STOC}, pages 1605--1616, 2024.
\newblock Revised version: arXiv:2404.17509, 2025.
\newblock \href {https://doi.org/10.1145/3618260.3649749}
  {\path{doi:10.1145/3618260.3649749}}.

\bibitem{cao-12-kernel-cluster-editing}
Yixin Cao and Jianer Chen.
\newblock Cluster editing: Kernelization based on edge cuts.
\newblock {\em Algorithmica}, 64(1):152--169, 2012.
\newblock \href {https://doi.org/10.1007/s00453-011-9595-1}
  {\path{doi:10.1007/s00453-011-9595-1}}.

\bibitem{charikar-05-approximate-clustering}
Moses Charikar, Venkatesan Guruswami, and Anthony Wirth.
\newblock Clustering with qualitative information.
\newblock {\em Journal of Computer and System Sciences}, 71(3):360--383, 2005.
\newblock \href {https://doi.org/10.1016/j.jcss.2004.10.012}
  {\path{doi:10.1016/j.jcss.2004.10.012}}.

\bibitem{chawla-06-hard-multicut}
Shuchi Chawla, Robert Krauthgamer, Ravi Kumar, Yuval Rabani, and D.~Sivakumar.
\newblock On the hardness of approximating multicut and sparsest-cut.
\newblock {\em Computational Complexity}, 15(2):94--114, 2006.
\newblock \href {https://doi.org/10.1007/S00037-006-0210-9}
  {\path{doi:10.1007/S00037-006-0210-9}}.

\bibitem{chawla-15-correlation-clustering}
Shuchi Chawla, Konstantin Makarychev, Tselil Schramm, and Grigory Yaroslavtsev.
\newblock Near optimal {LP} rounding algorithm for correlation clustering on complete and complete $k$-partite graphs.
\newblock In \emph{STOC}, pages 219--228, 2015.
\newblock \href {https://doi.org/10.1145/2746539.2746604}
  {\path{doi:10.1145/2746539.2746604}}.

\bibitem{cohen-addad-23-preclustering}
Vincent Cohen{-}Addad, Euiwoong Lee, Shi Li, and Alantha Newman.
\newblock Handling correlated rounding error via preclustering: {A}
  1.73-approximation for correlation clustering.
\newblock In \emph{FOCS}, pages 1082--1104, 2023.
\newblock \href {https://doi.org/10.1109/FOCS57990.2023.00065}
  {\path{doi:10.1109/FOCS57990.2023.00065}}.

\bibitem{demaine-06-correlation-clustering}
Erik~D. Demaine, Dotan Emanuel, Amos Fiat, and Nicole Immorlica.
\newblock Correlation clustering in general weighted graphs.
\newblock {\em Theoretical Computer Science}, 361(2-3):172--187, 2006.
\newblock \href {https://doi.org/10.1016/j.tcs.2006.05.008}
  {\path{doi:10.1016/j.tcs.2006.05.008}}.

\bibitem{dessmark-07-edge-clique-partition}
Anders Dessmark, Jesper Jansson, Andrzej Lingas, Eva{-}Marta Lundell, and Mia
  Persson.
\newblock On the approximability of maximum and minimum edge clique partition
  problems.
\newblock {\em International Journal of Foundations of Computer Science},
  18(2):217--226, 2007.
\newblock \href {https://doi.org/10.1142/S0129054107004656}
  {\path{doi:10.1142/S0129054107004656}}.

\bibitem{fan-26-pseudometric-weighted-correlation-clustering}
Chenglin Fan, Dahoon Lee, and Euiwoong Lee.
\newblock Pseudometric-weighted correlation clustering via spectral
  preclustering, 2026.
\newblock \href {https://doi.org/10.48550/arXiv.2609.17403}
  {\path{doi:10.48550/arXiv.2609.17403}}.

\bibitem{garcia-soriano-26-1.387}
David Garc\'{i}a-Soriano and Antoine Schohn.
\newblock Approximate dual separation for the cluster {LP}: a 1.387
  approximation for correlation clustering, 2026.
\newblock \href {https://doi.org/10.48550/ARXIV.2607.27829}
  {\path{doi:10.48550/ARXIV.2607.27829}}.

\bibitem{kalavas-25-constrained-correlation-clustering}
Andreas Kalavas, Evangelos Kipouridis, and Nithin Varma.
\newblock Towards better-than-2 approximation for constrained correlation
  clustering.
\newblock In \emph{ICML}, PMLR 267:28721--28734, 2025.
\newblock URL: \url{https://proceedings.mlr.press/v267/kalavas25a.html}.

\bibitem{khot-23-ugc}
Subhash Khot, Dor Minzer, and Muli Safra.
\newblock Pseudorandom sets in {G}rassmann graph have near-perfect expansion.
\newblock {\em Annals of mathematics}, 198(1):1--92, 2023.
\newblock \href {https://doi.org/10.4007/annals.2023.198.1.1}
  {\path{doi:10.4007/annals.2023.198.1.1}}.

\bibitem{khot-08-vc-ugc}
Subhash Khot and Oded Regev.
\newblock Vertex cover might be hard to approximate to within {$2 - \epsilon$}.
\newblock {\em Journal of Computer and System Sciences}, 74(3):335--349, 2008.
\newblock \href {https://doi.org/10.1016/j.jcss.2007.06.019}
  {\path{doi:10.1016/j.jcss.2007.06.019}}.

\bibitem{lee-25-chromatic-correlation-clustering}
Dahoon Lee, Chenglin Fan, and Euiwoong Lee.
\newblock 1.64-approximation for chromatic correlation clustering via chromatic cluster {LP}, 2026.
\newblock \href {https://doi.org/10.48550/arXiv.2507.15417}
  {\path{doi:10.48550/arXiv.2507.15417}}.

\bibitem{openai-26-ugc}
{OpenAI}.
\newblock The {U}nique {G}ames {T}heorem, 2026.
\newblock URL:
  \url{https://github.com/openai/math/tree/main/preprints/The-Unique-Games-Theorem-September-23-2026}.

\bibitem{puleo-15-bounded-cluster-sizes}
Gregory~J. Puleo and Olgica Milenkovic.
\newblock Correlation clustering with constrained cluster sizes and extended
  weights bounds.
\newblock {\em SIAM Journal on Optimization}, 25(3):1857--1872, 2015.
\newblock \href {https://doi.org/10.1137/140994198}
  {\path{doi:10.1137/140994198}}.

\bibitem{shamir-04-graph-modification-problem}
Ron Shamir, Roded Sharan, and Dekel Tsur.
\newblock Cluster graph modification problems.
\newblock {\em Discrete Applied Mathematics}, 144(1-2):173--182, 2004.
\newblock \href {https://doi.org/10.1016/j.dam.2004.01.007}
  {\path{doi:10.1016/j.dam.2004.01.007}}.

\bibitem{thomasse-10-kernel-fvs}
St{\'e}phan Thomass{\'e}.
\newblock A {$4 k^{2}$} kernel for feedback vertex set.
\newblock {\em ACM Transactions on Algorithms}, 6(2):32.1--32.8, 2010.
\newblock \href {https://doi.org/10.1145/1721837.1721848}
  {\path{doi:10.1145/1721837.1721848}}.

\bibitem{zuylen-09-deterministic-approximation}
Anke van Zuylen and David~P. Williamson.
\newblock Deterministic pivoting algorithms for constrained ranking and
  clustering problems.
\newblock {\em Mathematics of Operations Research}, 34(3):594--620, 2009.
\newblock \href {https://doi.org/10.1287/moor.1090.0385}
  {\path{doi:10.1287/moor.1090.0385}}.

\bibitem{veldt-22-clustering-via-strong-triadic-closure}
Nate Veldt.
\newblock Correlation clustering via strong triadic closure labeling: Fast approximation algorithms and practical lower bounds.
  \newblock In \emph{ICML}, PMLR 162:22060--22083, 2022.
  \newblock URL: \url{https://proceedings.mlr.press/v162/veldt22a.html}.
  
\bibitem{veldt-24-greedy-cover}
Nate Veldt.
\newblock Growing a random maximal independent set produces a 2-approximate
  vertex cover.
\newblock In \emph{SOSA}, pages 355--362, 2024.
\newblock \href {https://doi.org/10.1137/1.9781611977936.32} {\path{doi:10.1137/1.9781611977936.32}}.
\end{thebibliography}
\end{document}